\documentclass[aps,twocolumn,showpacs]{revtex4}

\usepackage{latexsym}
\usepackage{graphicx}
\usepackage{amsmath}

\begin{document}

\title{Fluctuation induced vortex pattern and its disordering
\\ in the fully frustrated $XY$ model on a dice lattice}

\author{S. E. Korshunov}
\affiliation{L. D. Landau Institute for Theoretical Physics,
Kosygina 2, Moscow 119334, Russia}

\date{October 27, 2004}

\begin{abstract}

A highly degenerate family of states, in which the plaquettes with
the same sign of vorticity form clusters of three [proposed in
Phys. Rev. B {\bf 63}, 134503 (2001)] is proven to really
minimize the Hamiltonian of the fully frustrated $XY$ model on a
dice lattice. The harmonic fluctuations are demonstrated to be of
no consequence for the removal of the accidental degeneracy of
these states, so a particular vortex pattern can be stabilized
only by the anharmonic fluctuations. The structure of this pattern
is found and the temperature of its disordering due to the
proliferation of domain walls is estimated. The extreme smallness
of the fluctuation induced free energy of domain walls leads to
the anomalous prominence of the finite-size effects, which prevents
the observation of vortex-pattern ordering in numerical
simulations. In such a situation the loss of phase coherence may
be related to the dissociation of pairs of fractional vortices
with the topological charges $\pm 1/8$. In a physical situation
the magnetic interactions of currents in a Josephson junction array 
will  be a more important source for the stabilization of a particular 
vortex pattern then the anharmonic fluctuations.

\end{abstract}


\maketitle

\section{Introduction}

The uniformly frustrated $XY$ model has been introduced by Teitel
and Jayaprakash \cite{TJ-L} for the description of a regular array
of superconducting islands connected with each other by Josephson
junctions (a Josephson junction array \cite{ML}) in the presence
of a uniform magnetic field. During the last two decades the main
attention has been concentrated on the investigation
\cite{TJ,LCJW,MS,ShS85,Hals,KU,K86,ffx-MC,FCKF,fxd,ffx,CF,fxh}
of so-called fully frustrated models
(on various lattices), which in terms of array correspond to having a
half-integer number of the superconducting flux quanta per plaquette
\cite{TJ}. The models belonging to this class can be also used for
the description of a planar magnet in which the neighboring spins
can have either ferromagnetic, or antiferromagnetic interaction,
the number of the antiferromagnetic bonds in each plaquette being
odd \cite{V77}.

The ground states of the uniformly frustrated $XY$ models are
characterized by the combination of the continuous and discrete
degeneracies \cite{TJ-L}. The former is related to the invariance
of energy with respect to the global phase rotation, whereas the
latter can be discussed in terms of the formation of a particular
vortex pattern.
In the fully frustrated models the numbers of plaquettes which 
contain positive and negative vortices should be equal to each other.

Since vortices of the same signs repel each other, the energy is
minimized when the vorticities of neighboring plaquettes are of
the opposite signs.
On a square lattice this requirement can be simultaneously satisfied
for all pairs of neighboring plaquettes, which allows one to conclude
that the ground state has the checkerboard structure
and a twofold discrete degeneracy \cite{V77}.
With increase of temperature
two different phase transitions can be expected to take place
\cite{TJ}, one of which is related to the loss of phase coherence
and the other can be associated with vortex-pattern disordering.
The analysis of the mutual influence between the two types of
topological excitations allows one to conclude that in the fully
frustrated $XY$ model on a square lattice the former has to take
place at lower temperature than the latter \cite{ffx}.

Triangular lattice also allows for the construction  of a
doubly degenerate pattern in which the vorticities are of the
opposite signs for all pairs of neighboring plaquettes
\cite{LCJW,MS}. It turns out possible to show that the thermodynamic
properties of the fully frustrated $XY$ model on a triangular lattice
(including the sequence of phase transitions) are completely
analogous to those of the model on a square lattice \cite{ffx}.

On a honeycomb lattice the situation is more complex, because
the family of ground states is characterized by an infinite
accidental degeneracy \cite{ShS85}, which can be described in terms
of the formation of parallel zero-energy domain walls \cite{K86}.
In such a case the structure of vortex pattern at low, but finite
temperatures cannot be determined without taking into account the
contribution to free energy from the small amplitude fluctuations in
the vicinities of different ground states.
This mechanism of the removal of an accidental degeneracy
\cite{VBCC,Shen} is often referred to as ``order-from disorder".
In systems with a continuous degeneracy it is usually sufficient to
compare the contributions from harmonic fluctuations \cite{Shen,Kaw,KVB}.

Recently it has been discovered  that in the fully frustrated $XY$
model on a honeycomb lattice the order-from-disorder mechanism
does not work at the harmonic level \cite{fxh}. The difference
between the free energies of fluctuations appears only when one takes
into account the anharmonicities, and, as a consequence, is
proportional not to the first, but to the second power of
temperature. This feature leads to the unusual prominence of the
finite-size effects \cite{fxh}.

\begin{figure}[b]
\includegraphics[width=60mm]{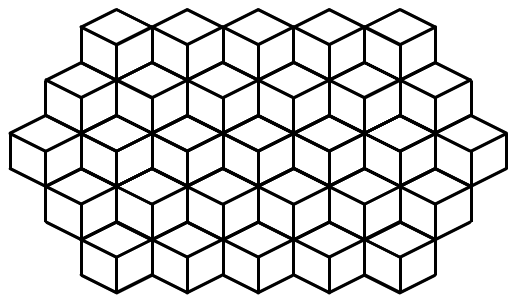}
\caption[Fig. 1] {Dice lattice is the simplest periodic lattice which
can be constructed from identical rhombic
plaquettes with three different orientations.}\label{fig1}
\end{figure}

The present work is devoted to the investigation of the fully
frustrated $XY$ model on  dice lattice \cite{HC} (see
Fig. 1). Like square, triangular and honeycomb lattices, dice
lattice consists of identical plaquettes (which in this case are
rhombic) and equivalent bonds. Since the
invariant description of a Josephson junction array can be
achieved only in terms of variables which are defined on lattice
bonds [the gauge-invariant phase differences, see Eq.
(\ref{theta})], and not on sites, dice lattice can be considered
as one of the four basic lattices for the investigation of the
uniformly frustrated $XY$ models. On more complex lattices
(containing nonequivalent plaquettes)  the correspondence
between an array and a frustrated $XY$ model is likely
to be broken as a consequence of the phenomenon of the ``hidden
incommensurability", related to the redistribution of
magnetic field between the plaquettes by screening currents in
asymmetric superconducting islands \cite{efc,PH}.

Recently a hypothesis has been but forward \cite{fxd} that in the
ground states of the fully frustrated $XY$ model on a dice lattice
the vortices of the same sign form three-vortex clusters (triads),
and a highly degenerate family of states has been proposed,
which satisfies this criterion and can be described in terms of
the formation of a network of intersecting zero-energy domain walls.
In Sec. II we present a rigorous prove that these states
indeed correspond to the absolute minimum of energy.

Sec. III is devoted to the analysis of harmonic fluctuations. We
reveal the existence of a hidden gauge symmetry,
%between the Hamiltonians of the harmonic fluctuations
% in the vicinity of different ground states,
which allows one to conclude that for a particular choice of
boundary conditions the set of the eigenvalues of the harmonic
Hamiltonian is exactly the same for all ground states. As a
consequence, the free energy of the harmonic fluctuations
% is the same in all ground states, and
cannot be the source for the selection of a particular vortex
pattern. This conclusion is valid also for quantum generalizations
of the model. All these properties resemble very much the
analogous properties of the fully frustrated $XY$ model on a
honeycomb lattice \cite{fxh}.

In Sec. IV   the lowest order %(in temperature)
contribution to the free energy of  anharmonic fluctuations is
considered. In particular, the gauge symmetry mentioned above is
applied to demonstrate that the contribution related to the
fourth-order terms in the Hamiltonian is the same for all ground
states and, therefore, is of no consequence for the selection of a
vortex pattern. The continuous approximation is used to show that
the fluctuation induced interaction of zero-energy domain walls is
extremely weak and decays inversely proportionally to the fifth
power of the distance between them. The comparison of the
numerically calculated free energies of anharmonic fluctuations in
different periodic ground states allows us to establish the vortex
pattern which can be expected to be stabilized at low
temperatures, and to find the fluctuation induced free energy of
zero-energy domain walls.

In Sec. V we analyze how the vortex pattern selected by anharmonic
fluctuations becomes disordered when one takes into account the
fluctuations of another type, namely, the formation of domain
walls, and propose an estimate for the temperature of the phase
transition which can be associated with the proliferation of such
defects. In this section we also discuss the finite-size effects
which interfere with the observation of vortex-pattern
ordering in finite samples and show that in the considered system
they are extremely prominent (exactly for the same reasons as in
the case of a honeycomb lattice).

In Sec. VI the interplay between the vortex-pattern disordering
and the loss of phase coherence is considered, whereas Sec. VII is
devoted to a discussion of another mechanism of the removal of an
accidental degeneracy related to magnetic interactions of currents
in the array \cite{PH,nwd}. In the concluding Sec. VIII our
results are summarized and compared with the results of numerical
simulations of Cataudella and Fazio \cite{CF}.

The interest to magnetically frustrated systems
with dice lattice geometry has appeared after Vidal {\em et al.}
\cite{VMD} have discovered that at full frustration
the ground state of a single electron which can jump between the
nearest sites of a dice lattice is infinitely degenerate and that
the corresponding wave function can be chosen as an arbitrary linear
combination of an infinite number of extremely localized wave
functions, each of which covers only a finite (and small) number
of sites \cite{VMD}.
Since it is known that the structure of the superconducting state in
a wire network is determined (in the mean-field approximation)
by the structure of the ground-state wave function of the
single-electron problem with the same geometry \cite{dG},
a suggestion has been put forward that at full frustrateion
the superconducting state in a dice network may have a
disordered (glassy-like) structure \cite{VMD,Abil}.
However, recently it has been shown \cite{nwd} that the inclusion into
analysis of the forth-order term of the Ginzburg-Landau functional
strongly decreases the ambiguity in the determination of the
structure of superconducting state in a fully frustrated wire
network with the dice lattice geometry. The set of states
minimizing the free energy of such a network turns out to be in
one-to-one correspondence with the set of the ground states of the
fully frustrated $XY$ model discussed on this article. In recent
years magnetically frustrated wire networks and Josephson junction
arrays with the dice lattice geometry have both been the subject
of active experimental investigations \cite{Abil,Pann,TTM}.

\section{The model and the ground states}
\subsection{The definition of the model}

A uniformly frustrated $XY$ model \cite{TJ-L} can be defined by
the Hamiltonian \cite{fn1}
\begin{equation}
H=-J\sum_{(jk)}\cos(\varphi_{k}-\varphi_{j}-A_{jk})\;,\label{H}
\end{equation}
where the summation is performed over all bonds $(jk)$ of a
regular two-dimensional lattice. In terms of a Josephson junction
array $J$ is the Josephson coupling constant of a single junction,
fluctuating variables $\varphi_{j}$ are the phases of the order
parameter on superconducting grains $j$ forming the array, whereas
quenched variables,
\begin{equation}                                  \label{Ajk}
 A_{jk}=\frac{2\pi}{\Phi_0}\int_{{\bf r}_j}^{{\bf r}_k}d{\bf r}\,
 {\bf A}({\bf r})\;,
\end{equation}
are defined by the integral of the vector potential ${\bf A}({\bf
r})$ of the external magnetic field along the bond $(jk)$,
$\Phi_0$ being the superconducting flux quantum. The form of Eq.
(\ref{H}) assumes that the currents in the array are sufficiently
small, so their proper magnetic fields can be neglected.

When the magnitude of the field corresponds to a half-integer
number of flux quanta per plaquette, the directed sum of
$A_{jk}\equiv -A_{kj}$ along the perimeter of a plaquette in
positive direction (which below is designated as $\sum_{\Box}$)
has to satisfy the constraint
\begin{equation}
\sum_{\Box}A_{jk}=\pi~(\mbox{mod }2\pi)          \label{SumA}
\end{equation}
on each plaquette of the lattice. In such a case the model is
called fully frustrated \cite{TJ}. In a more general case of a
uniformly frustrated $XY$ model, the right hand-side of Eq.
(\ref{SumA}) should be replaced by $2\pi f(\mbox{mod }2\pi)$,
where the frustration parameter $f$ describes the magnitude of the
external magnetic field in terms of the number of flux quanta per
plaquette. It is sufficient to consider the interval
\makebox{$f\in[0,\frac{1}{2}]$}, because all other values of $f$
can be reduced to this interval by a simple replacement of
variables \cite{TJ-L}. The term ``fully frustrated" is used for
the case of $f=1/2$, the maximal irreducible value of $f$.

Since both variables $\varphi_j$ and variables $A_{jk}$ depend on
a choice of a gauge, it is often more convenient to describe
different states of the system in terms of the gauge-invariant
phase differences,
\begin{equation}                                  \label{theta}
\theta_{jk}=\varphi_{k}-\varphi_{j}-A_{jk}\equiv -\theta_{kj}\;,
\end{equation}
defined on lattice bonds. Below we will always assume
$\theta_{jk}$ to be reduced to the interval $(-\pi,\pi)$. It
follows from the definition of these variables that in the fully
frustrated model they have to satisfy the constraints,
\begin{equation}
\sum_{\Box}\theta_{jk}=\pi~(\mbox{mod }2\pi)\;, \label{SumTh}
\end{equation}
completely analogous to Eq. (\ref{SumA}). One usually says that a
given plaquette contains a positive (or negative) half-vortex when
the left-hand side of Eq. (\ref{SumTh}) is equal to $+\pi$ (or
$-\pi$). Different minima of (\ref{H}) (including the ground
states) can be then identified in terms of a corresponding vortex
configuration.

The variation of Eq. (\ref{H}) with respect to $\varphi_{j}$
results in the current conservation equation for the site $j$, the
value of the current in the junction $(jk)$ being given by
\begin{equation}
I_{jk}=I_0\sin\theta_{jk}\;,                    \label{Ijk}
\end{equation}
where $I_0=(2e/\hbar)J$ is the critical current of a single
junction.

\subsection{Minimization of energy}

Dice lattice (Fig. 1)  is formed by two types of sites, with the
coordination numbers three and six, connected with each other in
such a way that each bond connects two sites with different
coordination numbers. Below we will always use index $k$ to denote
the threefold coordinated sites of a dice lattice and index $j$ to
denote the sixfold coordinated sites. For example, the bond $(jk)$
connects the sixfold coordinated site $j$ with the threefold
coordinated site $k$.

The minimization of the Hamiltonian (\ref{H}) with respect to all
variables $\varphi_k$ for the given values of the variables
$\varphi_j$ can be performed exactly. To describe the result of
this procedure it is convenient to introduce also the
gauge-invariant phase differences $\chi_{jj'}\equiv-\chi_{j'j}$
defined on the bonds of the triangular lattice $\cal T$ formed by
the sixfold coordinated sites $j$. A natural way to do it consists
in requiring that for each triangle formed by the sites $j$, $j'$
and $k$ (where $j$ and $j'$ are the nearest neighbors of $k$) the
sum of the three gauge-invariant phase differences taken along its
perimeter in the positive direction should be equal to $\pm\pi/2$
modulo $2\pi$, where the sign should be the same for all
triangles. In what follows we assume this sign to be negative,
\begin{equation}
\chi_{jj'}+\theta_{j'k}+\theta_{kj}=-\pi/2~(\mbox{mod }2\pi) \;.
                                             \label{SumTh2}
\end{equation}
Since $\chi_{jj'}=-\chi_{j'j}$, the constraint (\ref{SumTh}) then
automatically follows from Eq. (\ref{SumTh2}). On the other hand,
on each triangular plaquette of $\cal T$  the directed sum of the
variables $\chi_{jj'}$ has to satisfy the constraint
\begin{equation}
\sum_{\Box}\chi_{jj'}=\pi/2~(\mbox{mod }2\pi)\;, \label{SumChi}
\end{equation}
which can be obtained by summation of Eq. (\ref{SumTh2}) for three
neighboring triangles with the common cite $k$.

The minimization of
\[%begin{equation}
E_k=-J\sum_{a=1}^{3}\cos{\theta_{j_a k}}\;,
\]%end{equation}
(where $j_a$ with $a=1,2,3$ are the three nearest neighbors of $k$
on a dice lattice numbered in the positive direction) with respect
to $\varphi_k$ for the given values of $\chi_{j_1 j_2}$,
$\chi_{j_2 j_3}$ and $\chi_{j_3 j_1}$ satisfying the constraint
(\ref{SumChi}) gives
\[
E_k=-J\sqrt{3-2Y_k}
\]
where
\[
Y_{k}=%\sum_{a=1}^3
 \sin{\chi_{j_1 j_{2}}}+\sin{\chi_{j_2 j_{3}}}
 +\sin{\chi_{j_3 j_{1}}} \leq 3/2\;.
\]

Since $E(Y)=-J\sqrt{3-2Y}$ is a concave function of $Y$ and the
sum of the variables $Y_k$ over the whole lattice with the
periodic boundary conditions should be equal to zero, the absolute
minimum of
\[
H=\sum_k E(Y_k)
\]
on such a lattice is achieved when $Y_k=0$ for all $k$.
In the next subsection we demonstrate that this requirement
can be simultaneously satisfied on all plaquettes.
Accordingly, the value of energy in the absolute minimum is given by
$E=-2\sqrt{3}JN$, where $N$ is the total number of the sixfold
coordinated sites.

\subsection{Construction of ground states}

In the case of an isolated triangle the system of two equations
\begin{eqnarray*}
\chi_1+\chi_2+\chi_3 & = & \pi/2\;,          \\
\sin\chi_1+\sin\chi_2+\sin\chi_3 & = & 0\;,
\end{eqnarray*}
for three variables $\chi_1$, $\chi_2$ and $\chi_3$ has an
infinite number of solutions. However, the requirement to match
the solutions on all triangular plaquettes of $\cal T$ leads to
the removal of this continuous degeneracy.

\begin{figure}[b]
\includegraphics[width=68mm]{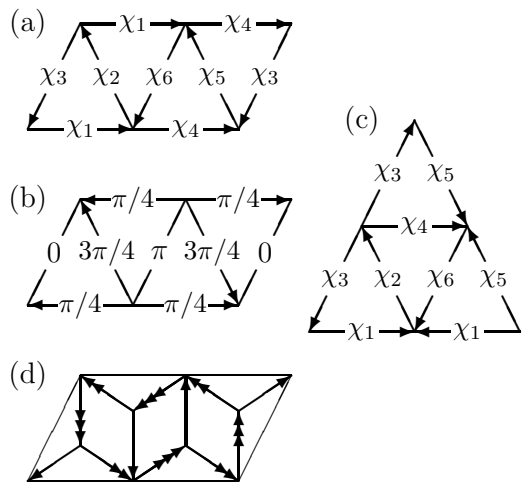}
\caption %[Fig. 2]
{Construction of a periodic ground state:
(a) a possible structure of an elementary cell;
(b) one of the solutions of Eqs. (\ref{SumChi3})-(\ref{SinChi3});
(c) an alternative elementary cell;
(d) the same elementary cell as in (b), but in terms of
$\theta_{jk}$.} %\label{fig2}
\end{figure}

Since the form of the constraint (\ref{SumChi}) corresponds to
having one quater of flux quantum per plaquette,
the minimal elementary cell whose periodic repetition allows to
construct a periodic solution consists of 4 triangles.
Such a solution can be described by 6 variables $\chi_i$
[defined, for example, like is shown in Fig. 2(a)], which have to
satisfy 3 constraints of the form (\ref{SumChi}):
\begin{subequations}                \label{SumChi3}
\begin{eqnarray}
 \chi_1+\chi_2+\chi_3 & = & \pi/2\;, \\
 \chi_4+\chi_5+\chi_6 & = & \pi/2\;, \\
-\chi_3-\chi_4-\chi_5 & = & \pi/2\;,
\end{eqnarray}
\end{subequations}
and 3 equations of the form $Y_k=0$:
\begin{subequations}                      \label{SinChi3}
\begin{eqnarray}
\sin\chi_1+\sin\chi_2+\sin\chi_3 & = & 0\;, \\
\sin\chi_4+\sin\chi_5+\sin\chi_6 & = & 0\;, \\
-\sin\chi_3-\sin\chi_4-\sin\chi_5 & = & 0\;,
\end{eqnarray}
\end{subequations}
The fourth constraint,
\[
-\chi_1-\chi_2-\chi_6 = -3\pi/2\;,
\]
and the fourth equation,
\[
-\sin\chi_1-\sin\chi_2-\sin\chi_6  = 0\;.
\]
are then satisfied automatically, as follows from the summation
of Eqs. (\ref{SumChi3}) and Eqs. (\ref{SinChi3}), respectively.

We have found that the numerical solution of the system of six
equations (\ref{SumChi3})-(\ref{SinChi3}) by an iterative
method always converges to the solution shown in Fig. 2(b)
or to other analogous solutions in which variables $\chi$ are
equal to  $0$, $-\pi/4$ and $3\pi/4$ on one half of triangular
plaquettes and to $\pi$, $\pi/4$ and $\-3\pi/4$ on the other half.
Note that each variable $\chi_{}$ belongs to two neigboring
plaquettes, but manifests itself on them with the opposite signs.
The same results are also obtained when one assumes that the
elementary cell has a different shape shown in Fig. 2(c).

Quite remarkably, an attempt to construct a solution with a
larger elementary cell leads to an overdefined system of equations.
For example, an elementary cell consisting of 8 triangles requires to
introduce 12 variables $\chi_i$ which have to satisfy
7 independent constraints of the form (\ref{SumChi})
and 7 independent equations of the form $Y_k=0$.
Apparently, one cannot expect a system of 14 equations for 12
variables to have a nontrivial solution which cannot be constructed
from the solutions obtained for a 4-triangle elementary cell.
Thus, the reduction of the problem to a triangular lattice has allowed
us to make a conclusion on the size of the elementary cell which
would be hardly possible in the framework of analysis in terms of
the original variables $\theta_{jk}$ defined on the bonds of a dice
lattice.

In Fig. 2(d) the structure of the elementary cell of Fig. 2(b) is
shown in terms of the variables $\theta_{jk}$.
Here single, double and triple arrows correspond, respectively,
to three different values of $\theta_{jk}$,
\[
\theta_{1,3} =
\arccos{\left(\frac{1}{\sqrt{3}}\pm\frac{1}{\sqrt{6}}\right)}\,,~~
\theta_{2} = \arccos\left({\frac{1}{\sqrt{3}}}\right)\,,
\]
satisfying the constraints
\[
\theta_2-\theta_1=\pi/4\;,~~ \theta_1+\theta_3=\pi/2\;,~~
\theta_2+\theta_3=3\pi/4\;,
\]
which lead to the fulfilment of Eq. (\ref{SumTh}) on all rhombic
plaquettes, and the current conservation equation,
\[
  \sin\theta_1+\sin\theta_2=\sin\theta_3\;,
\]
the form of which follows from Eq. (\ref{Ijk}).

The most compact way of illustrating a structure of a given state
(a local or global minimum of the Hamiltonian) consists in showing
which plaquettes are occupied by positive and which plaquettes by
negative half-vortices. In Fig. 3(a) this approach is used to
demonstrate the structure of the ground state which is obtained by
the periodic repetition of the elementary cell shown in Fig. 2(d).
Notice that positive and negative half-vortices are
grouped into clusters of three (triads).
The rules which allow one to restore the values of $\theta_{jk}$ for
each bond from the structure of vortex pattern (in a ground state)
can be found in Ref. \onlinecite{fxd}.

\begin{figure}[b]
\includegraphics[width=85mm]{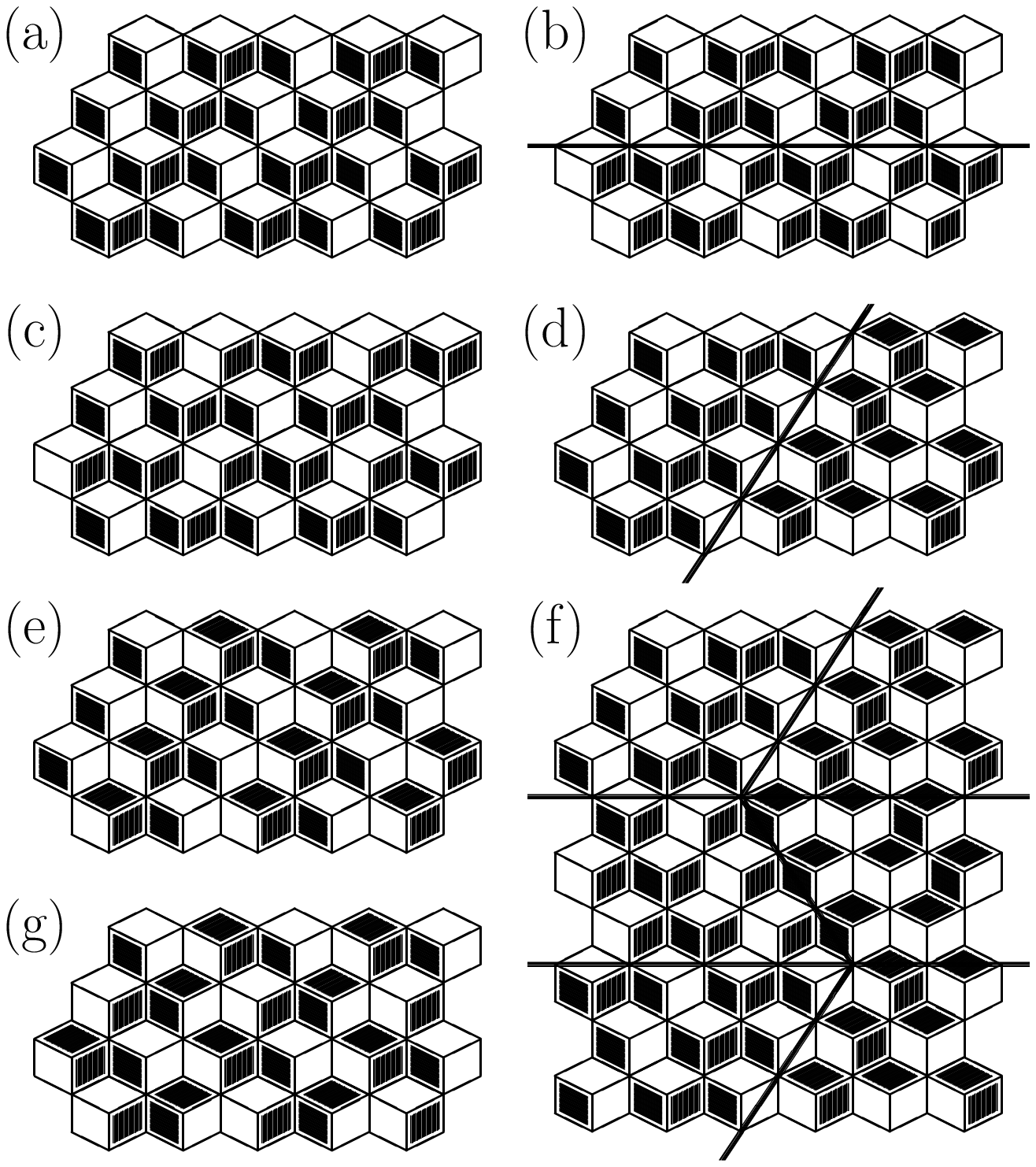}
\caption[Fig. 3]{ The structure of some ground states.
The plaquettes with positive vorticities are marked in black.
}\label{fig3}
\end{figure}

Since positive and negative half-vortices can be considered as
occuping the cites of the dual lattice (which in the present case is
a {\em kagome} lattice), the structure of a given state of the
fully frustrated $XY$ model on a dice lattice can be compared with the
structures of different states of the antiferromagnetic Ising model
on a {\em kagome} lattice.
In particular, according to Wolf and Schotte \cite{WS}, in the
framework of the Ising model the state with the structure shown in
Fig. 3(a) is selected when \makebox{$\overline{J}_1\gg
\overline{J}_2\gg \overline{J}_4>0$} and all other couplings are equal
to zero. Here $\overline{J}_i$ is the coupling constant for $i$th
neighbors on a {\em kagome} lattice, and we have used the notation of
Ref. \onlinecite{nwd} for the classification of dice lattice
plaquettes as neighbors of each other.

The ground state whose structure is shown in Figs. 2(d) and 3(a)
allows for creation of infinite domain walls which
brake the periodicity of this state but does not cost any energy
\cite{fxd}. These zero-energy domain walls can be of the two types.

Fig. 3(b) shows an example of zero-energy domain wall of the type
I. Note that the orientations of triads formed by negative
half-vortices (white plaquettes) are different above and below the
wall.
The configuration of arrows (defining the values of the variables
$\theta_{jk}$) after crossing such a wall can be obtained from the
old configuration (in the absence of the wall) by its reflection
with respect to a line which is perpendicular to the wall and
subsequent inversion of all arrows. There can exist an arbitrary
number of such domain walls in parallel to each other. By creating
them at every possible position one obtains another periodic
ground state shown in Fig. 3(c).

Fig. 3(d) shows an example of zero-energy domain wall of the type
II. After crossing such a wall the variables $\theta$ are changed
(in comparison with what they would be in the absence of the wall)
according to even more simple rule, which can be codified as
\begin{equation}
\theta_1\rightarrow\theta_3\,,~\theta_3\rightarrow\theta_1\,,~
\theta_2\rightarrow -\theta_2\,.                    \label{ruleII}
\end{equation}
Note that both black and white triads have different orientations
on two sides of the wall, whereas at the wall the shape of white
triads is modified.

Again, there can exist an arbitrary number of the type II domain
walls parallel to each other. By inserting them at every possible
position one obtains one more periodic ground state shown in Fig.
3(e), in which all triads have the modified shape. Alternatively,
one can start the whole construction from the periodic state of
Fig. 3(e) and obtain the state of Fig. 3(a) by introducing a dense
sequence of domain walls on crossing which the same rule, Eq.
(\ref{ruleII}), is applicable.

The zero-energy domain walls of different types can cross each
other without increasing energy. However, it follows from
the rule for the transformation of the state induced by the type I
domain wall (described above) that a type II domain wall should
change its orientation by $\pi/3$ each time it crosses a type I
domain wall [see Fig. 3(f)]. A dense network of zero-energy domain
walls of both types constructed on the background of state (a)
leads to the periodic ground state shown in Fig. 3(g). The
structures of the periodic states (a), (c), (e) and (g) in terms
of the variables $\theta_{jk}$ are shown in Fig. 3 of Ref.
\onlinecite{nwd}.

Note that the formation of zero-energy domain walls is related to
the changes of the orientation of vortex triads (and, in the case
of type II domain walls, also of their shapes), but does not lead
to the appearance of vortex clusters of other sizes.

\section{Harmonic approximation}
\subsection{Two families of eigenmodes}

The Hamiltonian describing the harmonic fluctuations in the
vicinity of one of the ground states described in Sec. II.C can be
written as
\begin{equation}
H^{(2)}=\frac{1}{2}\sum_{(jk)}J_{jk}(u_j-v_k)^2       \label{H2}
\end{equation}
where $u_{j}$ are deviations of the variables $\varphi_j$ from
their equilibrium values on sixfold coordinated sites, $v_{k}$ are
analogous deviations on threefold coordinated sites and the
coupling constants $J_{jk}\equiv J\cos\theta_{jk}$ acquire one of
the three possible values
\begin{equation}                            \label{J123}
J_{1,3}=\left(\frac{1}{\sqrt{3}}\pm\frac{1}{\sqrt{6}}\right)J\,,~~
J_2=\frac{J}{\sqrt{3}}\,,
%~~ J_3=\left(\frac{1}{\sqrt{3}}-\frac{1}{\sqrt{6}}\right)J\,,
\end{equation}
in accordance with the value of $\theta_{jk}$ on the bond $(jk)$.

If phase dynamics in a Josephson junction array can be assumed to
be non-dissipative and the capacitance matrix of the array has
only diagonal elements, the linearized equations of motion for the
variables $v_k$ and $u_j$ following from Eq. (\ref{H2}) can be
written as
\begin{subequations}                    \label{leuv}
\begin{eqnarray}
(2J_S-M_6\omega^2)u_{j} & = & \sum_{k=k(j)}J_{jk}v_k\;,
                                          \label{leu}\\
(J_S-M_3\omega^2)v_{k} & = & \sum_{j=j(k)}J_{jk}u_j\;,
                                            \label{lev}
\end{eqnarray}
\end{subequations}
where $J_S=J_1+J_2+J_3$, $M_i=(\hbar/2e)^2C_i$ (where $\,i=3,6$),
and $j(k)$ denotes the nearest neighbors of $k$. In Eqs.
(\ref{leuv}) we have performed the Fourrier transformation to the
frequency representation and have assumed that the
self-capacitances of the superconducting islands, $C_3$ and $C_6$,
are different for the two types of islands.

Note that in all ground states discussed in Sec. II.C the coupling
constants $J_{jk}$ always have the same three values ($J_1$, $J_2$
and $J_3$) on the three bonds $(j_a k)$ connected to any site $k$,
as a consequence of which the coefficient standing in the
left-hand side of Eq. (\ref{lev}) does not depend on $k$. This
allows to conclude that all eigenmodes with \makebox{$u_j\equiv
0$} should have the same eigenfrequency
$\omega_0=({J_S/M_3})^{1/2}$. In the thermodynamic limit the
degeneracy of this eigenfrequency is equal to one third of the
total number of modes.
% whereas in a finite system with open boundary formed by only six-fold
% coordinated sites it is given by the difference between the
% numbers of threefold and sixfold coordinated sites.

The spectrum of the brunch with $u_{j}\neq 0$ can be found from
the equation which is obtained after substitution of Eq.
(\ref{lev}) into Eq. (\ref{leu}) and can be written as
\begin{equation}                                  \label{leu2}
\Lambda(\omega)u_{j}=\sum_{j'=j'(j)} {K}_{jj'} (u_j-u_{j'})\;,
\end{equation}
where
\begin{equation}
\Lambda(\omega)=(2M_3+M_6)\omega^2-\frac{M_3 M_6}{J_S}\omega^4\;,
\end{equation}
$j'(j)$ are the six nearest neighbors of $j$ on $\cal T$ and
\begin{equation}                            \label{Kjj}
K_{jj'}={(J_{jk'}J_{j'k'}+J_{jk''}J_{j'k''})}/{J_S}\;,
\end{equation}
$k'$ and $k''$ being the two threefold coordinated sites
belonging to the same rhombic plaquette as $j$ and $j'$.

The right-hand side of Eq. (\ref{leu2}) has exactly the same form
as in the equation describing harmonic fluctuations on a
triangular lattice with the nearest neighbor interaction
characterized by the coupling constants ${K}_{jj'}$ defined by Eq.
(\ref{Kjj}). Quite remarkably, in all the ground states described
above these coupling constants acquire only two values
\begin{subequations}                    \label{K12}
\begin{eqnarray}
K_1 & = & \frac{2J_1 J_3}{J_S}=\frac{J}{3\sqrt{3}}\;, \label{K1}\\
K_2 & = & \frac{(J_1+J_3)J_2}{J_S}=\frac{2J}{3\sqrt{3}}
\label{K2}\;,
\end{eqnarray}
\end{subequations}
which differ from each other by the factor of 2. When a
half-vortex in the plaquette $\langle jk'j'k''\rangle$ is the
central vortex of a triad to which it belongs, $K_{jj'}=K_1$,
whereas otherwise $K_{jj'}=K_2$. In all the ground states
described in Sec. II.C these couplings are distributed between the
bonds of $\cal{T}$ in such a way that in each  triangular
plaquettes one bond has $K_{jj'}=K_1$, whereas the two other bonds
have $K_{jj'}=K_2$.

\subsection{Comparison of different ground states}

\begin{figure}[b]
\includegraphics[width=60mm]{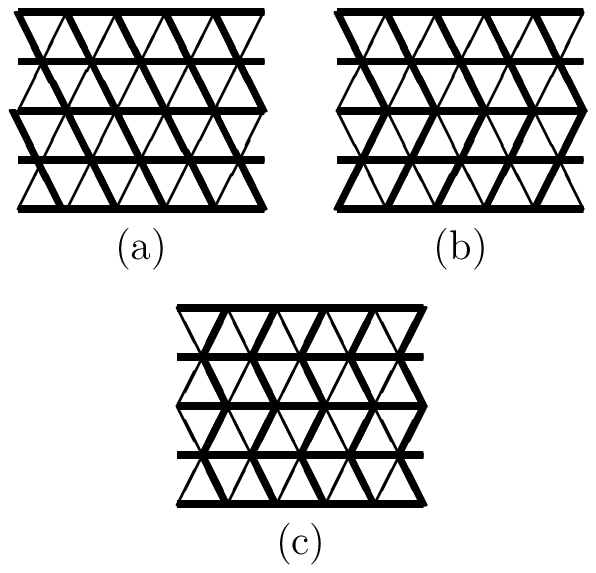}
\caption[Fig. 4] {The distribution of coupling constants $K_{jj'}$
between the bonds of $\cal T$ in different states: (a) states $a$
and $e$; (b) a single type I domain wall; (c) states $c$ and $g$.
Thin lines correspond to $K_{jj'}=K_1$ and thick lines to
$K_{jj'}=K_2$. \label{fig4}}
\end{figure}

\begin{figure}[b]
\includegraphics[width=60mm]{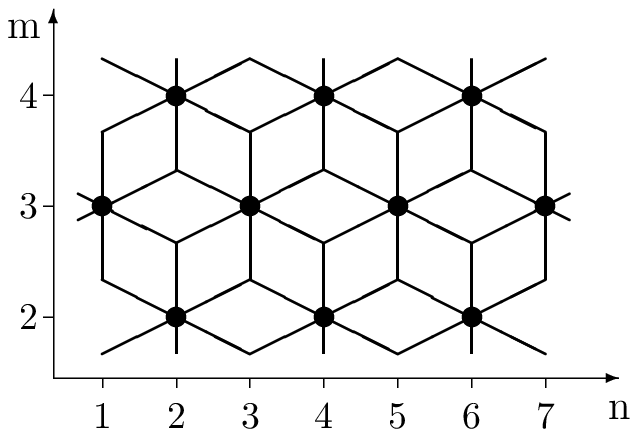}
\caption[Fig. 5] {The numbering of the sixfold coordinated sites
by pairs of integers $(n,m)$ with the same parity. \label{fig5}}
\end{figure}

In the state (e) all sixfold coordinated sites have the identical
environment in terms of the coupling constants $J_{jk}$. As a
consequence, the values of the coupling constants $K_{jj'}$ in
this state depend only on the orientation of the bonds $(jj')$.
For two of the three possible orientations of the bonds they are
equal to each other, see Fig. 4(a). If all sixfold coordinated
sites $j$ are renumbered by pairs of integers with the same parity
$(n,m)$ as shown in Fig. 5, Eq. (\ref{leu2}) in this state can be
rewritten as
\begin{equation}                                 \label{leu3}
\begin{array}{rcl}
[2K_S & - & \Lambda(\omega)]u_{n,m}=                    \\
& = & {K}_{1}^{(m)}u_{n+1,m+1}+{K}_2^{(m)} u_{n-1,m+1}   \\
& + & {K}_1^{(m-1)}u_{n-1,m-1}  +  {K}_2^{(m-1)}u_{n+1,m-1}  \\
& + & {K}_2[u_{n+2,m}+u_{n-2,m}]\;,
\end{array}
\end{equation}
where $K_S=K_1+2K_2$ and
\begin{equation}                            \label{Km}
{K}_1^{(m)}={K}_1\;,~~{K}_2^{(m)}={K}_2\;,
\end{equation}
for all values of $m$.

For $K_{1,2}^{(m)}$ given by Eq. (\ref{Km}), Eq. (\ref{leu3}) can
be easily solved after performing the Fourrier transformation.
Substitution of \makebox{$u_{n,m}\propto \exp i(qn+pm)$} into Eq.
(\ref{leu3}) gives the dispersion relation in the form
\begin{eqnarray}                              \label{disp-rel}
%\begin{array}{rcl}
\Lambda(\omega)& = & 2K_S   -  2K_1\cos(q+p)  \nonumber      \\ &
- & 2K_2[\cos(q-p)+\cos 2q]\;.
% \end{array}
\end{eqnarray}
Note that Eq. (\ref{disp-rel}) is of the second order in
$\omega^2$, which corresponds to the existence of the two
momentum-dependent eigenfrequencies, $\omega_1(q,p)$ and
$\omega_2(q,p)$, for each point in the Brillouin zone in addition
to \makebox{$\omega_0=({J_S/M_3})^{1/2}$} discussed above.

According to Eq. (\ref{ruleII}), each zero-energy domain wall of
the type II leads to the permutation of the coupling constants
$J_1$ and $J_3$ in the Hamiltonian of harmonic fluctuations. Such
a permutation does not change the factor $J_S-M_3\omega^2$ in the
right-hand side of Eq. (\ref{lev}), and therefore does not change
neither the degeneracy, nor the frequency,  $\omega_0$, of the
family of the eigenmodes with $u=0$. Since Eqs. (\ref{K12}) are
also invariant with respect to the permutation of $J_1$ and $J_3$,
such a permutation introduces no changes to the form of Eq.
(\ref{leu2}) as well. This means that the whole set of the
eigenfrequencies of the harmonic fluctuations in the system does
not feel the presence of the type II domain walls and is exactly
the same in all the states which can be obtained from each other
by the insertion of the type II domain walls. For example, the
dispersion relation (\ref{disp-rel}) is valid not only in the
state (e), but also in the state (a), which is characterized by
exactly the same pattern of the coupling constants $K_{jj'}$ shown
in Fig. 4(a).

The zero-energy domain walls of the type I lead to a more complex
permutation of coupling constants. However, in the terms of the
coupling constants $K_{jj'}$ the consequences of this permutation
are rather simple and reduce to the permutation of ${K}_1$ and
${K}_2$ for those two orientations of the bonds $(jj')$ that are
not parallel to the direction of the wall \cite{footnote3}, see
Fig. 4(b). This means that Eqs. (\ref{Km}) should be valid only
when $d_m$, the number of the type I domain walls situated at
$m'\leq m$, is even and should be replaced by
\[
{K}_1^{(m)}={K}_2\;,~~{K}_2^{(m)}={K}_1\;
\]
when $d_m$ is odd.

In the presence of periodic boundary conditions in the horizontal
direction and open boundary conditions in the perpendicular
(vertical) direction the irrelevance of such permutations of
coupling constants for the set of eigenfrequencies can be easily
demonstrated in the framework of the mixed representation which is
obtained after performing the Fourrier transformation with respect
to the variable $n$, keeping the variable $m$ as it is. In the
terms of the variable $u_m(q)$ defined in such a way, Eq.
(\ref{leu3}) can be rewritten as
\begin{eqnarray}                             \label{leu4}
[2K_S & - & \Lambda(\omega)]u_m(q) ={K}^{(m)}{(q)}u_{m+1}(q) \\
& + & 2{K}_2\cos(2q)u_m(q)+ {K}^{(m-1)}(q)u_{m-1}(q)\;,\nonumber
\end{eqnarray}
where
\[
{K}^{(m)}{(q)} = K_0(q)\exp[i\alpha(q) s_m]\;.
\]
The dependence of ${K}^{(m)}{(q)}$ on $m$ enters only through
$s_m=(-1)^{d_m}=\pm 1$, whereas both
\[
K_0(q) = |{K}_1\exp(iq)+{K}_2\exp(-iq)|\;
\]
and
\[
\alpha(q) = \arg[{K}_1\exp(iq)+{K}_2\exp(-iq)]\;
\]
are independent of $m$.

With the help of the simple gauge transformation
\begin{equation}                                  \label{gauge-tr}
u_m(q)=\exp\left[i\alpha(q)\sum_{m'<m}s_m\right]{u}'_m(q)\;,
\end{equation}
which cannot not change the eigenvalues, Eq. (\ref{leu4}) can be
transformed to the form
\begin{equation}                                 \label{leu5}
\begin{array}{rcl}
[2K_S & - & \Lambda(\omega)]{u}'_m(q)=K_0(q){u}'_{m+1}(q) \\
 & + & 2{K}_2\cos(2q){u}'_m(q)+ K_0(q){u}'_{m-1}(q)\;,
\end{array}
\end{equation}
in which all coefficients do not depend on $m$. This property of
Eq. (\ref{leu5}) proves that for the considered boundary
conditions the whole set of eigenfrequencies is insensitive to the
presence of domain walls of the type I. Since Eq. (\ref{leu5}) has
been derived from Eq. (\ref{leu3}), the form of which does not
depend on the presence of the zero-energy domain walls of the type
II, the same conclusion is applicable also in the presence of an
intersecting network of zero-energy domain walls of both types.

Note that this does not mean that the spectrum of fluctuations in
an infinite system, understood as the dependence of the
eigenfrequencies on  the two-dimensional wave-vector, is the same
for all periodic ground states. On the contrary, it should  have
different forms in the states whose transformation into each other
requires the insertion of a periodic sequence of the type I domain
walls. In particular, the dispersion relation in the state (c) and
state (g) [which can be obtained from each other by the insertion
of the dense sequence of type II domain walls and both are
characterized by the distribution of the coupling constants
$K_{jj'}$ shown in Fig. 4(c)] can be obtained by substitution of
$u_m(q)\propto \exp(ipm)$ into Eq. (\ref{leu5}) and is of the form
\begin{equation}
\Lambda(\omega) = 2K_S-2K_0(q)\cos p-2K_2\cos 2q\;,
                                   \label{disp-rel-c}
\end{equation}
Apparently, it does not coincide with the dispersion relation in
the states (a) and (e) given by Eq. (\ref{disp-rel}). Note that
the elementary cell of the state (c) consists of twelve sites,
and, therefore, a more straightforward approach to the derivation
of its dispersion relation would give it in the form of the
determinant of a twelve by twelve matrix.

However, the free energy of harmonic fluctuations, $F_2$, which in
a general situation (that is, when the quantum effects are also
taken into account) can be written as
\begin{equation}
F_2=T\sum_{\{\omega\}}\ln\left(2\sinh\frac{\hbar\omega}{2
T}\right)\;                                       \label{F2}
\end{equation}
is determined entirely  by the set of the eigenfrequencies of the
system $\{\omega\}$. Thus, our results demonstrate that for the
considered boundary conditions the value of $F_2$ is exactly the
same for all the ground states discussed above even in the case of
a finite system. For other types of boundary conditions the same
property will be recovered in the thermodynamic limit.

Naturally, these conclusions should remain valid both in the
zero-temperature limit, when Eq. (\ref{F2}) is transformed into
the expression for the energy of zero-point fluctuations,
\[%begin{equation}
E_2=\frac{\hbar}{2}\int\int\frac{dq\,dp}{(2\pi)^2}
\left[\omega_0+\omega_1(q,p)+\omega_2(q,p)\right]\;,
\]%end{equation}
and in the classical limit ($\hbar\rightarrow 0$), when the
dynamical properties of the system are of no importance, and Eq.
(\ref{F2}) is reduced to
\[%begin{equation}                                \label{F2i}
F_2=%\mbox{const}+
\frac{T}{2}\int\int\frac{dq\,dp}{(2\pi)^2}%\frac{dp}{2\pi}
\ln\left[\frac{\Lambda(q,p)}{T}\right]\;,
\]%end{equation}
where $\Lambda(q,p)$ denotes the function of $q$ and $p$ standing
in the right-hand side of the corresponding dispersion relation,
Eq. (\ref{disp-rel}) or Eq.(\ref{disp-rel-c}).

Another system, in which the accidental degeneracy of its ground
states remains unbroken when the free energy of the harmonic
fluctuations is taken into account, is the fully frustrated $XY$
model with a honeycomb lattice \cite{fxh}. The method used in this
section (the construction of the gauge transformation which
reduces the linearized equations of motion for fluctuations in
different ground states to the same form) presents a
generalization of the approach of Ref. \onlinecite{fxh}, where the
analogous gauge transformation has been constructed for the
harmonic part of the Hamiltonian.

The analysis of this  section can be generalized for the case
when the capacitance matrix of a Josephson junction array in
addition to the self-capacitances of superconducting islands also
takes into account $C_1$, the capacitances of Josephson junctions
forming the array. This will lead to the replacement
\[J_a\rightarrow {J}_a(\omega)\equiv J_a-M_1\omega^2\;,\]
where $M_1=(\hbar/2e)^2C_1$, in all dispersion relations, but will
not bring any changes in the qualitative conclusions. It is
equally possible to include into consideration an Ohmic
dissipation of each junction described by a frequency-dependent
harmonic contribution to its Euclidean action.

\subsection{Correlation functions}

Since the evolution of the variables $u_j$ can be described by
Eqs. (\ref{leu2}), which contain only coupling constants
$K_{jj'}$, but not the original coupling constants $J_{jk}$, the
symmetry of the correlation function of the variables $u_j$ in a
given state will be determined entirely by the symmetry of the
configuration of $K_{jj'}$ in this state. In particular, in
situation when the value of $K_{jj'}$ depends only on the
orientation of the bond $(jj')$ and can be equal only to $K_1$ or
$K_2$, the correlation function
\begin{equation}
G_{jj'}=\langle(u_{j}-{u_{j'}})^2\rangle\;,      \label{Gjj}
\end{equation}
where $j$ and $j'$ are the nearest neighbors of each other on
$\cal{T}$, should also be dependent only on the orientation of the
bond $(jj')$ and acquire one of the two possible values, which
below are denoted $G_1$ and $G_2$. The same property can described
by saying that in such states $G_{jj'}$ depends only on whether
$K_{jj'}$ is equal to $K_1$ or to $K_2$,
\begin{equation}
G_{jj'}=\left\{\begin{array}{ll}
G_1~~ & \mbox{for }K_{jj'}=K_1\;, \\
G_2~~ & \mbox{for }K_{jj'}=K_2\;.
\end{array}\right.                       \label{G12}
\end{equation}

The type II domain walls are simply of no consequence for the
values of $K_{jj'}$ and, therefore, for the correlation functions
of the variables $u$. On the other hand, it is rather evident that
the gauge transformation (\ref{gauge-tr}) leaves invariant the
form of
the correlation functions % $\langle(u_{n,m}-{u_{n',m}})^2\rangle$
of the variables $u_{n,m}$ with the same value of $m$. This means
that  the correlation function $\langle(u_{j}-{u_{j'}})^2\rangle$
cannot be sensitive to the presence of the type I domain walls
which do not pass between the points $j$ and $j'$. Since the sites
$j$ and $j'$, which are the nearest neighbors of each other on
$\cal T$, are too close to have a domain wall passing between
them, Eq. (\ref{G12}) will be applicable also in the presence of
an arbitrary set of zero-energy domain walls.

\section{Anharmonic fluctuations}
\subsection{Basic relations}

In this section and below our analysis is restricted to the
classical version of the model. It is well known that in a
classical system the leading contribution to the free energy
related to anharmonic fluctuations, \makebox{$F_{\rm
anh}=F_3+F_4$}, is the sum of the two terms,
\begin{equation}
F_3=-\frac{1}{2T}\left\langle
\left[H^{(3)}\right]^2\right\rangle\; \label{F3}
\end{equation}
and
\begin{equation}
F_4=\langle H^{(4)}\rangle \;,                         \label{F4}
\end{equation}
where $H^{(3)}$ and $H^{(4)}$ are, respectively, the third- and
the forth-order contributions to the expansion of the Hamiltonian
in the vicinity of a particular ground state, and angular brackets
denote the averages over thermodynamic fluctuations  calculated
with the help of the harmonic Hamiltonian.

In the considered problem $H^{(3)}$ and $H^{(4)}$ can be written
as
\[%begin{equation}                              \label{H3}
H^{(3)}=\sum_{k}H^{(3)}_k;~~ H^{(3)}_k=
-\frac{1}{6}\sum_{j=j(k)}J'_{jk}(u_j-v_k)^3\;
\]%end{equation}
and
\[%begin{equation}                               \label{H4}
H^{(4)}=\sum_{k}H^{(4)}_k;~~ H^{(4)}_k=
-\frac{1}{24}\sum_{j=j(k)}J_{jk}(u_j-v_k)^4\;,
\]%end{equation}
where coupling constants $J_{jk}$ are exactly the same as in the
harmonic Hamiltonian, Eq. (\ref{H2}), whereas coupling constants
$J'_{jk}=J\sin\theta_{jk}$ acquire one of the six possible values
$J'_{jk}=\pm J\sin\theta_a$ in accordance with the value of
$\theta_{jk}$ on the bond $(jk)$. Due to the current conservation
condition coupling constants $J'_{jk}$ have to satisfy the
constraints,
\begin{equation}                        \label{SumJj}
\sum_{j=j(k)}J'^{}_{jk}=0\;,~~
%\end{equation} and \begin{equation}    \label{SumJk}
\sum_{k=k(j)}J'_{jk}=0\;,
\end{equation}
on all sites of the lattice.

\subsection{Invariance of $F_4$}

In the case of a dice lattice there exists a convenient way to
separate the fluctuations on the two types of sites from each
other, which allows to considerably simplify the calculation of
the averages in Eqs. (\ref{F3}) and (\ref{F4}). It consists in the
replacement of variables
\begin{equation}
v_k=w_k+v^{(0)}_k\;,~                             \label{w}
v^{(0)}_k={\sum_{j=j(k)}J_{jk}u_j}/J_S\;,
\end{equation}
which transforms the harmonic Hamiltonian (\ref{H2}) into
\begin{equation}
H^{(2)}=\frac{1}{2}\sum_{(jj')}{K}_{jj'}(u_j-u_{j'})^2
+\frac{1}{2}\sum_{k}J_S w_k^2\;,                  \label{H2w}
\end{equation}
where the coupling constants $K_{jj'}$ are, naturally, the same as
have been obtained in Sec. III.B, see Eqs. (\ref{Kjj}) and
(\ref{K12}), after the exclusion of the variables $v_k$ from the
equations of motion.

A simple form of Eq. (\ref{H2w}), in which each variable $w_k$ is
decoupled from all other variables, makes the calculation of
averages with respect to the fluctuations of $w_k$ a very
straightforward procedure. Substitution of Eq. (\ref{w}) into the
expression for $H^{(4)}_k$ with subsequent expansion of the result
in powers of $w_k$ allows one to express the result of such an
averaging of $H^{(4)}_k$ as a fourth-order polynomial of the
variables $u_j$,
\begin{eqnarray}                              \label{P4}
&&P_4(u_{j_1},u_{j_2},u_{j_3}) =\\
&&-\frac{1}{24}\sum_{a=1}^3 J_{a} \left(3\langle w^2\rangle^2
+6\langle w^2\rangle \tilde{u}_{j_a}^2+ \tilde{u}_{j_a}^4\right)
\;,                                            \nonumber
\end{eqnarray}
where $\langle w^2\rangle={T}/{J_S}$ is the value of $\langle w_k
\rangle^2$, which is the same for all $k$, whereas
\[
  \tilde{u}_{j_a}\equiv u_{j_a}-v_k^{(0)}=
  \frac{1}{J_S}\sum_{b\neq a}J_b(u_{j_a}-u_{j_b})\;.
\]
All terms which are odd in $w_k$ have disappeared from Eq.
(\ref{P4}) due to the corresponding symmetry of the Hamiltonian
(\ref{H2w}). Note that the form of $P_4(u_1,u_2,u_3)$ does not
depend on $k$. To achieve that we have renumbered in Eq.
(\ref{P4}) the three sites $j$ which are the nearest neighbors of
$k$ as $j_a\equiv j_a(k)$ in such a way that $J_{j_a k}=J_a$.

Since Hamiltonian (\ref{H2w}) does not include linear terms, the
result  of the Gaussian averaging of
$P_4(u_{j_1},u_{j_2},u_{j_3})$ will have a form  of a  second
order polynomial, $P_2(G_{j_1j_2}, G_{j_2j_3}, G_{j_1j_3})$, whose
three  arguments are the  nearest neighbor correlation functions
defined by Eq. (\ref{Gjj}). Instead of looking for the explicit
form of $P_2$, it is sufficient to notice that since the central
vortex of a triad is always surrounded by the bonds with $J_{jk}$
equal to $J_1$ or $J_3$, we will always have $K_{j_1j_3}=K_1$ and
$K_{j_1j_2}=K_{j_2j_3}=K_2$, and, as a consequence of Eq.
(\ref{G12}), the result of the averaging of $H^{(4)}$ will have
the same form,
\[\langle H^{(4)}_k\rangle=P_2(G_2,G_2,G_1)\;,\]
for all $k$ independently of what particular ground state is
considered. Therefore, the value of $F_4=\sum_{k}\langle
H^{(4)}_k\rangle$ will be exactly the same for all the ground
states which we are trying to compare already at the level of the
separate terms in this sum.

\subsection{Simplification of $F_3$}

The result of the averaging of $H^{(3)}_k$ with respect to
fluctuations of $w_k$ can be in analogous way reduced to the sum
of two terms, the first of which,
\[%begin{equation}                                  \label{P1}
P_1(u_{j_1},u_{j_2},u_{j_3}) =-\frac{\langle w^2\rangle}{2}
\sum_{a=1}^{3} J'_{j_a k} \tilde{u}_{j_a}\;,
\]%end{equation}
is a first-order and the second,
\[%begin{equation}                                  \label{P3}
P_3(u_{j_1},u_{j_2},u_{j_3}) =-\frac{1}{6}\sum_{a=1}^{3}J'_{j_ak}
\tilde{u}_{j_a}^3\;,
\]%end{equation}
a third-order polynomial of the variables $u_{j_a}$, which are
assumed here to be numbered in the same way as in Eq. (\ref{P4}).
Both $P_1$ and $P_3$ depend on $k$ only through the factor
$\tau_k=\pm 1$, which, for example, can be chosen to be determined
by the sign of $J'_{j_1 k}$.

The sum of $P_1(u_{j_1},u_{j_2},u_{j_3})$ over $k$ can be
reordered as a sum over $j$,
%\begin{widetext}
\[
\sum_{k}\left.P_1(u_{j_1},u_{j_2},u_{j_3})\right|_k
=-\frac{\langle w^2\rangle}{2J_S}\sum_{j}C_j u_j\;,
\]
%\end{widetext}
where
\[
C_j=\left.\sum_{k=k(j)}
[J'_{jk}(J_{j'k}+J_{j''k})-J_{jk}(J'_{j'k}+J'_{j''k})] \right.
\]
and $j'$ and $j''$ are the other two nearest neighbors of $k$ in
addition to $j$. It is not hard to notice that all coefficients
$C_j$ are equal to zero as a consequence of Eqs. (\ref{SumJj}).

This  allows one to express the result of the averaging of
$\left[H^{(3)}\right]^2$ over fluctuations of the variables $w_k$
as
\begin{equation}                              \label{P6}
\left\langle\left[H^{(3)}\right]^2\right\rangle_w= R^2+
\sum_{k}P_6\left(u_{j_1},u_{j_2},u_{j_3}\right)
\end{equation}
where
\begin{equation}                              \label{R}
  R=\sum_{k}P_3\left(u_{j_1},u_{j_2},u_{j_3}\right)\;
\end{equation}
and $P_6$ is a sixth-order polynomial of its arguments, the form
of which is the same for all $k$. Exactly like it happens with
$P_4$, the averaging of $P_6\left(u_{j_1},u_{j_2},u_{j_3}\right)$
over fluctuations of $u_j$ produces the same expression for all
$k$ independently of what particular ground state is considered.
That means that any difference between the values of $F_{\rm anh}$
in different ground states can result only from the first term in
Eq. (\ref{P6}),
\begin{equation}\label{Fanh}
  F_{\rm anh}=\mbox{const}-\frac{1}{2T}\langle R^2\rangle\;.
\end{equation}

\subsection{Explicit expressions}

Let us start with comparing the free energies of anharmonic
fluctuations in the states (a) and (e). Instead of calculating
$F_{\rm anh}$ separately for both these states it is more
convenient to construct an explicit expression directly for
$\delta F^{\rm e,a}_{\rm anh}=F_{\rm anh}^{\rm e}-F_{\rm anh}^{\rm
a}$, the difference in $F_{\rm anh}$ between the states (e) and
(a). Since both these states are characterized by the same form of
the effective Hamiltonian for the variables $u_j$, the
construction of such an expression does not require the
application of the gauge transformation given by Eq.
(\ref{gauge-tr}). The same is true also for the whole set of
states which can be constructed from the state (a) or state (e) by
the insertion of some sequence of type II domain walls.

If the state (e) shown in Fig. 3(e) is rotated by $\pi/3$ in such
a way that the orientation of the possible type II domain walls
becomes horizontal, the expression for $R$,  Eq. (\ref{R}), in
this state can be rewritten as
\begin{equation}                             \label{Re}
  R_{\rm e}=\sum_{m} \sigma_m S_m^{+}\;,
\end{equation}
where we have again used the numbering of sites defined by Fig. 5,
$\sigma_m=(-1)^m$,
\begin{eqnarray}                             \label{Smnu}
  S_m^{\mu} &
  = & \sum_{n=m\,(\mbox{\scriptsize mod}\,2)}
  [P_3^{\mu}(u_{n+2,m},u_{n+1,m+1},u_{n,m})     \nonumber  \\
   & + & P_3^{\mu}(u_{n+1,m+1},u_{n,m},u_{n-1,m+1})]\;,
\end{eqnarray}
and $P_3^{+}=\tau_k P_3$ is an invariant version of $P_3$,
\begin{equation}                                  \label{P3+}
P_3^{+}(u_{1},u_{2},u_{3}) =-\frac{1}{6J_S^3}\sum_{a=1}^{3}J'_a
\left[\sum_{b\neq a}{J_b}(u_{a}-u_{b})\right]^3\;,
\end{equation}
the form of which does not depend on $k$. Instead of introducing a
definition simply for $S^+_m$, we have used Eq. (\ref{Smnu}) to
define a more general object $S_m^\mu$, where superscript $\mu$
can be equal to $\pm 1$ or $0$. However, for the compactness of
notation we will usually replace $\mu=\pm 1$ simply by plus or
minus. In Eq. (\ref{P3+}) the constants $J'_{jk}$ are expressed in
terms of three constants,
\begin{subequations}                             \label{J'123}
\begin{eqnarray}
J'_1 & = & J\sin\theta_1 =
\left(\frac{1}{\sqrt{3}}-\frac{1}{\sqrt{6}}\right)J\,,\\
J'_2 & = & J\sin\theta_2 = \frac{2}{\sqrt{6}}J\,,
\\
J'_3 & = & -J\sin\theta_3 =
-\left(\frac{1}{\sqrt{3}}+\frac{1}{\sqrt{6}}\right)J\,,
\end{eqnarray}
\end{subequations}
whose sum is equal to zero in accordance with Eqs. (\ref{SumJj}).

Substitution of Eqs. (\ref{J123}) and Eqs. (\ref{J'123}) into Eq.
(\ref{P3+}) allows to reduce it to
\begin{eqnarray}                         \label{P3+b}
  P_3^{+}(u_1,u_2,u_3) & = &
   K_3(u_2-u_3)^2(u_3-u_1) \\
   & + & K'_3(u_1-u_2)^3+K''_3(u_3-u_1)^3\;, \nonumber
\end{eqnarray}
where
\[
    K_3  =  \frac{J}{6\sqrt{6}}\,,~~
    K'_3 =  \frac{J}{9\sqrt{6}}\,,~~
    K''_3=
    \left(\frac{1}{9\sqrt{3}}-\frac{1}{36\sqrt{6}}\right)J\,.
\]
However, in all the ground states that we consider the last term
from Eq. (\ref{P3+b}) after substitution of $P_3=\tau_k P_3^+$
into Eq. (\ref{R}) cancels with analogous term from the
neighboring triangular plaquette, which allows one to put
$K_3''=0$.

Each time a type II domain wall is crossed the replacement of
variables $\theta_j$ described by Eqs. (\ref{ruleII}) has to take
place. In terms of the expression for $R$ this procedure is
translated into the replacement of each term of the form
$P_3^{+}(u_1,u_2,u_3)$ by
\[%begin{equation}                              \label{P3-}
P_3^{-}(u_1,u_2,u_3)=-P_3^{+}(u_3,u_2,u_1)\;.
\]%end{equation}
As a consequence, the value of $R$ for the state which is obtained
after the insertion of the dense sequence of type II domain walls
(the structure of this state is obtained after rotating Fig. 3(a)
by $\pi/3$) will have the form
\begin{equation}                             \label{Ra}
  R^{\rm a}=\sum_{m} \sigma_m S_m^{\sigma_m}\;.
\end{equation}

Substitution of Eqs. (\ref{Re}) and (\ref{Ra}) into Eq.
(\ref{Fanh}) allows to express $\delta F^{\rm e,a}_{\rm anh}$
%$=F_{\rm anh}^e-F_{\rm anh}^a$
as
\begin{equation}                        \label{Fea}
\delta F_{\rm anh}^{\rm e,a}=N\sum_{l=1}^{\infty} V({2l-1})\;,
\end{equation}
where $V(m))$ is the average
\begin{equation}                               \label{Vm}
V(m)=\frac{2}{TL_x}\left\langle S_{m_1}^0 S_{m_2}^0
\right\rangle\;,
\end{equation}
which depends only on $m=m_1-m_2$, $L_x$ is the size of the system
in the horizontal direction (in lattice unites of $\cal{T}$) and
$S_m^{0}$ is given by Eq. (\ref{Smnu}) with
\[
\begin{array}{lcr}                %          \label{P30}
  P_3^{0}(u_1,u_2,u_3)
  & = & \frac{1}{2}\left[P_3^{+}(u_1,u_2,u_3)
  -P_3^{-}(u_1,u_2,u_3)\right]           \\
  $~$ \\
  & = & \frac{1}{2}\left[P_3^{+}(u_1,u_2,u_3)
  +P_3^{+}(u_3,u_2,u_1)\right]
\end{array}
\]
being the symmetric (with respect to the permutation of $u_1$ and
$u_3$) part of $P_3^{+}(u_1,u_2,u_3)$. The explicit expression for
$P_3^{0}(u_1,u_2,u_3)$ which follows from Eq. (\ref{P3+b}) is
\begin{eqnarray}                             \label{P30b}
  P_3^{0}(u_1,u_2,u_3) & = &
   \frac{K_3}{2}(u_3-u_1)^2(u_3-2u_2+u_1) \nonumber      \\
   & + & \frac{K'_3}{2}[(u_1-u_2)^3+(u_3-u_2)^3]\;.
\end{eqnarray}

Analogous comparison of the values of $F_{\rm anh}$ in two
different states allows to find that the free energy of a single
type II domain wall on the background of the state (a) is given by
\begin{equation}                     \label{FDW}
 F_{\rm DW}=L_x\sum_{m=1}^{\infty}m V(m)\;,
\end{equation}
whereas the interaction of two domain walls situated at $m=m_1$
and $m=m_2$ can be written as
\begin{equation}                     \label{Fint}
  F_{\rm int}(m_2-m_1)=-2L_x\sum_{m=1}^{\infty}mV({|m_2-m_1|+m})\;.
\end{equation}

\subsection{Continuous approximation}

When Eq. (\ref{P30b}) is substituted in the expression for $S_m^0$
given by  Eq. (\ref{Smnu}) with $\mu=0$, all terms proportional to
$K'_3$ cancel each other after the summation over $n$, so it
becomes possible to rewrite this expression as
\begin{eqnarray}                               \label{Sm0}
  S_m^0 & = & 4K_3\left[\sum_{n=m'(\mbox{\scriptsize
               mod}\,2)}\right.\left.            \nonumber
            (\nabla_n u)^2(\nabla_{m'}^{+}u)\right|_{m'=m+1} \\
        &   & -\left.\sum_{n=m'(\mbox{\scriptsize mod}\,2)}(\nabla_n u)^2
  \left(\nabla_{m'}^{-}u)\right|_{m'=m}\right]\;,
\end{eqnarray}
where
\[%begin{equation}                            \label{Dnu}
  \nabla_n u\equiv \frac{u_{n+2,m'}-u_{n,m'}}{2}
\]%end{equation}
is a lattice analog of the derivative in the horizontal direction
and
\[%begin{equation}                            \label{Dmu}
  \nabla_{m'}^{\pm} u\equiv
  \pm\left[\frac{u_{n,m'}+u_{n+2,m'}}{2}-u_{n+1,m'\mp 1}\right]
\]%end{equation}
are two lattice analogs of the derivative in the vertical
direction suitable for a triangular lattice. It follows from
symmetry reasons that
\begin{equation}                           \label{DnuDmu}
  \left\langle(\nabla_{n}^{}u)(\nabla_{m}^{\pm}u)\right\rangle=0\;.
\end{equation}

The function $S_m^0$ defined by Eq. (\ref{Sm0}) is a third order
polynomial of variables $u_{n,m'}$ belonging to the stripe with
$m'=m,m+1$. Accordingly, the result of the Gaussian averaging of
the product $S^0_{m_1}S^0_{m_2}$ will  be a third order polynomial
of the two-point correlation functions. It is not hard to check
that the only terms which survive in the expression
for $\langle S_{m_1}S_{m_2}\rangle$ are the triple
products of the two-point correlation functions whose arguments
belong to different stripes, whereas all other terms cancel each
other in the result of the summation over $n$ or as a consequence
of Eq. (\ref{DnuDmu}). Speaking more precisely, due to the
structure of the expression for $S_m^0$, Eq. (\ref{Sm0}), they
are the triple products of the lattice analogs of the
derivatives of such correlation functions, which for $|m_1-m_2|\gg
1$ can be rather accurately calculated in the framework of the
continuous approximation.

When integer variables $n$ and $m$ are replaced by continuous
variables $x$ and $y$, both the first and the second term in the
square brackets in Eq. (\ref{Sm0}) should be replaced by the same
integral
\[%begin{equation}                            \label{Sm0a}
  \frac{1}{2}\int_{}^{} dx\, u_x^2 u_{y}\;,
\]%end{equation}
where, however, $u^2_x$ should be calculated at the values of $y$
which differ from each other by one (in this subsection subscripts
$x$ and $y$ designate partial derivatives with respect to the
corresponding variables). This means that in the framework of the
continuous approximation the total expression for $S_m^0$ should
be replaced by
\begin{equation}                            \label{Sm0b}
  (S_m^0)_{\rm cont}
  =\sqrt{2}K_3\int_{}^{} dx \,(u_x^2)_y u_y
  =-\sqrt{2}K_3\int_{}^{} dx \,u_{xx}u_y^2 \;.
\end{equation}
When writing Eq. (\ref{Sm0b}) we have performed the integration by
parts and also the rescaling $x\rightarrow \lambda x$
($\lambda^2=1+2K_1/K_2=2$), which transforms the continuous
version,
\[%begin{equation}                        \label{H2c}
  H^{(2)}_{\rm cont}=\frac{1}{2}\int_{}^{} \int_{}^{}dx\,
  dy\,\left[(2K_1+K_2)u_x^2+K_2 u_y^2\right]\;,
\]%end{equation}
of the harmonic Hamiltonian,
\begin{eqnarray*}
  H^{(2)} = \sum_{n=m(\mbox{\scriptsize mod}\,2)}
  \sum_{m} \left[\frac{K_1}{2}(u_{n+2,m}-u_{n,m})^2\right. & + & \\
   +\left.\frac{K_2}{2}\sum_{l=\pm 1}(u_{n+1,m+l}-u_{n,m})^2\right] & &\;,
\end{eqnarray*}
to the isotropic form,
\begin{equation}                        \label{H2ci}
  H^{(2)}_{\rm isotr}=\frac{K_{\rm eff}}{2}\int_{}^{} \int_{}^{}
  dx\,dy\, \left[u_x^2+u_y^2\right]\;,
\end{equation}
where
\[%begin{equation}                                  \label{Keff}
  K_{\rm eff}=\sqrt{(2K_1+K_2)K_2}
  =\left(\frac{2}{3}\right)^{3/2}J\;.
\]%end{equation}

Substitution of the correlation function,
\[%begin{equation}                            \label{G}
  G(x,y)=\mbox{const}-\frac{T}{2\pi K_{\rm eff}}
  \ln(x^2+y^2)^\frac{1}{2}\;,
\]%end{equation}
corresponding to the Hamiltonian (\ref{H2ci}) into
\begin{eqnarray*}                            \label{SS}
 & & \langle (S_{m_1}^0)_{\rm cont}(S_{m_2}^0)_{\rm cont}\rangle=\\
 & & = 2K_3^2\int_{-\infty}^{\infty}dx_1\int_{-\infty}^{\infty}dx_2
  [2G_{xxxx}G_{yy}^2+4G_{xxy}^2G_{yy}]\;,
\end{eqnarray*}
[where all derivatives of $G(x,y)$ should be taken at $x=x_1-x_2$,
$y=y_1-y_2$] and subsequent integration over $x_1-x_2$ give
\begin{equation}                             \label{Vmc}
  V_{\rm cont}(m)= \gamma_c\frac{T^2}{J}\frac{1}{|m_1-m_2|^7}\;,
\end{equation}
where
\[
  \gamma_c=2\sqrt{2}\frac{2K_3^2 J}{(2\pi K_{\rm eff})^3}
  \frac{15\pi}{4}=\frac{45\sqrt{3}}{1024\pi^2}\approx
  0.0077\;.
\]

Thus we have found that the quantity $V(m)$, summation of which
over $m$ allows to find different essential free energies,
contains a very small numerical coefficient $\gamma_c$ and very
rapidly decays with the increase of $m$. Accordingly, the
expressions in Eq. (\ref{Fea}) and Eq. (\ref{FDW}) which include
the summation of $V(m)$ starting from $m=1$ will be determined
entirely by the first term in the sum. However, substitution of
Eq. (\ref{Vmc}) into Eq. (\ref{Fint}) with subsequent replacement
of the summation by integration allows one to find the form of the
interaction of two domain walls for $m\gg 1$,
\[%begin{equation}\label{Fint2}
  F_{\rm int}(m)\approx-\frac{\gamma_c}{15}\frac{T^2}{J}\frac{L_x}{m^5}\;.
\]%end{equation}

\subsection{Numerical calculations}

Note that the expression for $V(m)$, Eq. (\ref{Vmc}), which we
have found in the framework of the continuous approximation is
positive and increases with decrease of $m$, that is when one
moves out of the region of the applicability of the continuous
approximation. Although it hardly can be expected that the more
accurate calculation will lead to the change of the sign of
$V(m)$, we have checked this for $m=1$ by going beyond the limits
of the continuous approximation.

The exact expression for $V(m)$ given by Eq. (\ref{Vm}) and Eq.
(\ref{Sm0}) can be written as a sum over all possible pairs of
triangles belonging to two different stripes. For $m=1$ the first
term in this sum, that is the contribution which corresponds to
the pair of adjacent triangles has the form
\begin{equation}                                  \label{V1p}
\begin{array}{rcl}
  V_0(1) & = & \frac{4K_3^2}{T}
  \langle(u_{n+2,m}-u_{n,m})^2\rangle^2
  \langle(\nabla^+_m u)(-\nabla^-_m u)\rangle\;,
\end{array}
\end{equation}
The averages which enter Eq. (\ref{V1p}) are given by the
integrals over Brillouin zone,
\begin{eqnarray}
\langle(u_{n+2,m}-u_{n,m})^2\rangle & = &
 \int_{-\pi}^{\pi}\int_{-\pi}^{\pi} \frac{dq\,dp}{(2\pi)^2}
 \frac{W_1(q,p)\,T}{\Lambda_0(q,p)}
%\approx 0.433\,\frac{T}{K_2}
\;,\nonumber\\
 \langle(\nabla^+_m u)(-\nabla^-_m u)\rangle & = &
 \int_{-\pi}^{\pi}\int_{-\pi}^{\pi} \frac{dq\,dp}{(2\pi)^2}
\frac {W_2(q,p)\,T} {\Lambda_0(q,p)}
%\approx 0.0294\,\frac{T}{K_2}
\;,\nonumber
\end{eqnarray}
where
\begin{eqnarray*}
W_1(q,p) & = & 2(1-\cos 2q)\;, \\
W_2(q,p) & = & (\cos q-\cos p)^2-\sin^2 p\;, \\
\Lambda_0(q,p) & = & 2K_1(1-\cos\,2q)+4K_2(1-\cos q \cos p)\;.
\end{eqnarray*}
Numerical calculation of these integrals gives
\begin{eqnarray*}                        \label{averages}
\langle(u_{n+2,m}-u_{n,m})^2\rangle & \approx &
0.433\,\frac{T}{K_2} \;,~~~ \\
 \langle(\nabla^+_m u)(-\nabla^-_m u)\rangle
 &  \approx & 0.0294\,\frac{T}{K_2} \;,
\end{eqnarray*}
which after substitution into Eq. (\ref{V1p}) allows one to find
that $V_0{(1)}=\gamma_{0}{T^2}/{J}$, where $\gamma_{0}\approx
0.0018$.

Addition to Eq. (\ref{V1p}) of the analogous terms related to the
pairs of triangular plaquettes which have a common site leads to
the replacement of $\gamma_{0}$ by $\gamma_2\approx 0.0032>0$. The
contributions from more distant pairs of plaquettes are much
smaller and can be safely neglected. This result confirms the
positiveness of $V(1)$. According to Eq. (\ref{Fea}), the
positiveness of $V(m)$ for all $m$ ensures that \makebox{$F_{\rm
anh}^{\rm e}>F_{\rm anh}^{\rm a}$}.

It follows from Eq. (\ref{FDW}) that the value of the type II
domain wall free energy per unit length,
%$\epsilon_{\rm DW}(T)=F_{\rm DW}(T)/L_x$,
can be then rather accurately estimated as
\begin{equation}                               \label{fDW0}
  f^{(0)}_{\rm DW}= \gamma_2\frac{T^2}{J}\;.
\end{equation}
In the analysis of the next section $f^{(0)}_{\rm DW}(T)$
plays the role of the fluctuation induced effective energy of a
domain wall.

The comparison of the free energies of anharmonic fluctuations in
the states (a) and (c) can be made following the same approach,
but turns out to be much more cumbersome for two reasons. First,
in order to reduce the Hamiltonians of the harmonic fluctuations
in these two states to the same form one has to apply the gauge
transformation introduced in Sec. III.B. Second, there is no
complete cancellation of the second term from Eq. (\ref{P3+b}),
which leads to the strong increase of the number of terms one has
to take into account in the expressions for the free energies of
fluctuations. A numerical calculation shows that the free energy
of anharmonic fluctuations is lower for the state (a), which means
that the free energy of the type I domain wall shown in Fig. 3(b)
is also positive.

The numerical constant  $\gamma_1$ characterizing this free energy
% ($\epsilon'_{\rm DW}=\gamma T^2/T$)
is equal to $0.0044$ if one takes into account only the
contributions from the adjacent plaquettes, whereas when the
contributions from the pairs of plaquettes which have a common
site are also included, one gets the value which is very close to
$\gamma_2$, $$\gamma_1\approx 0.0033$$.

\section{Disordering of vortex pattern}
\subsection{An estimate for the phase transition temperature}

The temperature of the phase transition associated with the
proliferation of domain walls and the disordering of the periodic
vortex pattern can be estimated by analyzing a more complete
expression for the domain wall free energy, $f_{\rm DW}(T)$, which
in addition to the term induced by anharmonicities, $f^{(0)}_{\rm
DW}(T)=\gamma T^2/J$, should also include the entropic term
related to the formation of kinks,
\[%begin{equation}                           \label{fDW}
  f_{\rm DW}(T)=f^{(0)}_{\rm DW}(T)-\nu T\exp(-E_{\rm K}/T)\;,
\]%end{equation}
where $E_{\rm K}\propto J$ is the energy of a kink and $\nu\sim 1$
is the density (per unit length) of the positions on a domain wall
where a kink can exist. In the case of the exactly solvable
anisotropic Ising model \cite{Ons44} an analogous estimate allows
one to find the transition temperature with the exponential
accuracy.

The temperature of the phase transition associated with the
spontaneous creation of infinite domain walls, $T_c$,
can be then estimated from the condition \makebox{$f_{\rm
DW}(T_c)=0$}, which can be rewritten as
\begin{equation}                            \label{Tc}
T_c=\frac{E_{\rm K}}{\ln\left[\nu T_c/f^{(0)}_{\rm DW}(T_c)\right]}\;.
\end{equation}
Eq. (\ref{Tc}) shows that $T_c$ is determined mainly by $E_{\rm K}$
and only logarithmically depends on $f^{(0)}_{\rm DW}$, that is on
$\gamma$.

\begin{figure}[b]
\includegraphics[width=65mm]{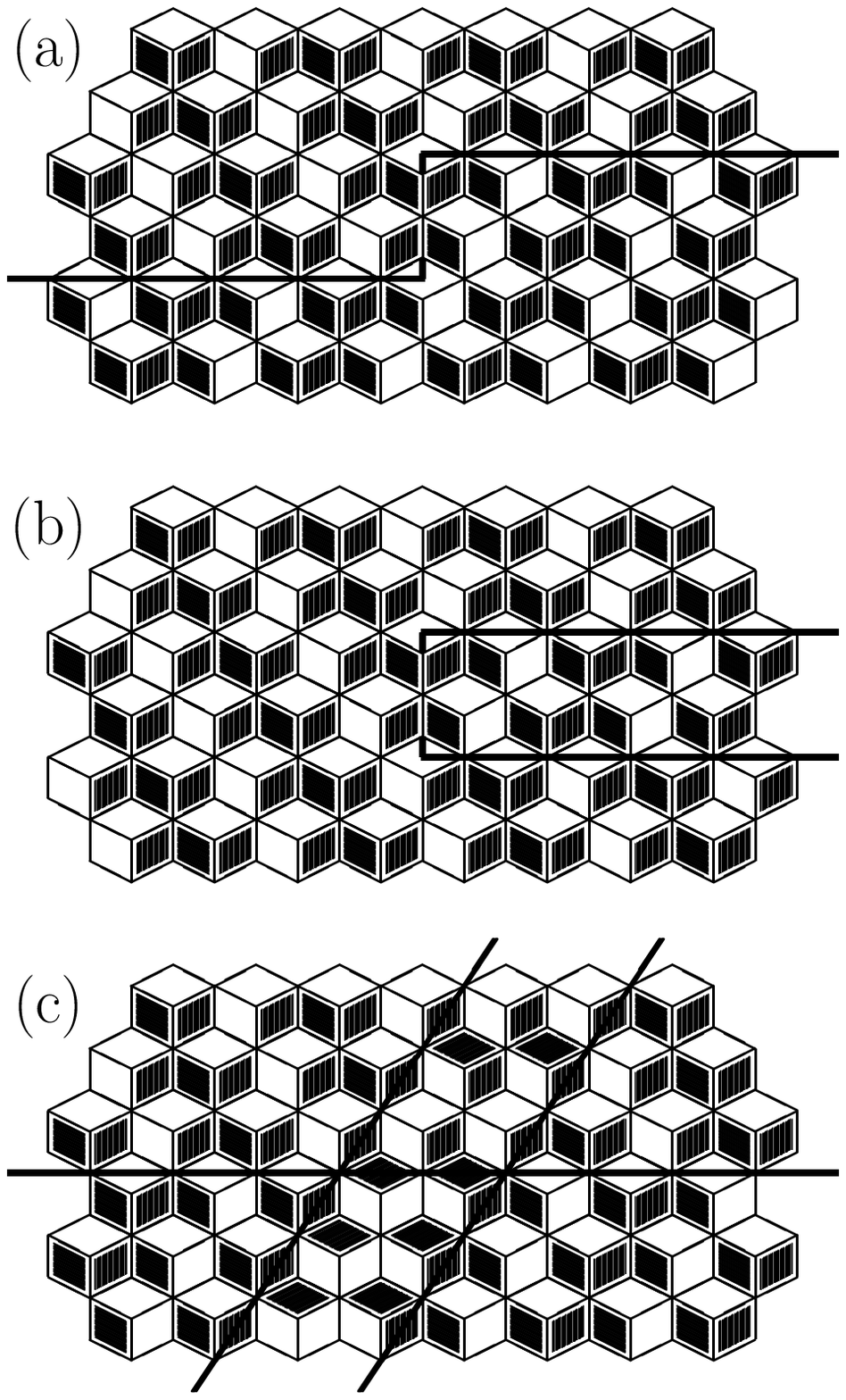}
\caption[Fig. 6] { Point-like defects with finite energies: (a) a
kink on a type I domain wall, (b) an end point of a striped defect
with the minimal width, (c) an intersection of a striped defect
with a type I domain wall.}\label{fig6}
\end{figure}

Fig. 6(a) shows the structure of an elementary kink on a type I
domain wall separating two different versions of the state (a).
This is one of the simplest neutral point-like excitations
possible in the system. It contains only two vortex clusters of
anomalous sizes, one with four positive vortices (instead of three)
and another with two, so there is no excess vorticity associated
with this defect. Note that any defect with only one vortex
cluster of anomalous size will be characterized by a non-zero
vorticity, so its energy will be logarithmically divergent.

The distance between two neighboring positions the kink shown in
Fig. 6(a) can occupy is equal to 2 (in lattice units of $\cal T$).
However, the same domain wall allows also for the formation
of kinks of the opposite sign (orientation), which means that the
value of $\nu$ in Eq. (\ref{Tc}) should be set equal to one.

Numerical calculation of the kink energy has been performed by
minimizing the energy of a finite lattice cluster around the kink
(with the size $4L\times 4L$, containing $N=48L^2-10L+1$ sites
inside it) with the assumption that on all sites outside of this
area the values of the phases are exactly the same as they would
be if the kink was infinitely far. Numerical calculation of
$E_{\rm K}$ for $L=1,2,3$ ($L=3$ corresponds to $N=403$) and
extrapolation of the result to $L\rightarrow\infty$ give
\begin{equation}                                 \label{EK}
  \frac{E_{\rm K}}{J}= 0.1037\pm 0.0001 \;.
\end{equation}

The simplest kink on a type II domain wall is also neutral, but has a
more complex structure. It contains four vortex clusters of anomalous
sizes, and, therefore, its energy is, roughly speaking,  two times
larger then the energy of a kink on a type I domain wall. This
means that in the vicinity of the phase transition type I domain
walls play a relatively more important role. Numerical solution of
Eq. (\ref{Tc}) for $E_{\rm K}/J=0.1037$ and $\gamma=0.0033$ gives
\begin{equation}                               \label{Tc-est}
T_c/J\approx 0.010\,.
\end{equation}

\subsection{Vortex pattern fluctuations in the low temperature phase}

At $T<T_c$ all domain walls excited as thermal fluctuations should
have the form of closed loops. At temperatures well below $T_c$
the form of these loops will be strongly anisotropic. At $T\ll
T_c$ a typical defect will be formed by two parallel zero-energy
domain walls separated by the minimal possible distance \cite{KVB}.
Accordingly, the free energy of such stripe defect can be written
as
\[%begin{equation}                            \label{E_D}
  F_{\rm SD}(L,T)=2E_0+2f^{(0)}_{\rm DW} (T) L\;,
\]%end{equation}
where $E_0$ is the energy which can be associated with each of the
two ends of stripe defect and $L$ is its length.
Since in the considered problem
the structure of the end point of a striped defect is the same as the
one of the kink [see Fig. 6(b)], $E_0$ is very close to $E_{\rm K}$.

The probability of formation of stripe defects, $P_{\rm SD}(L)$,
is, naturally, determined by their free energies,
\[%begin{equation}                          \label{pDL}
  P_{\rm SD}(L)=\exp[-F_{\rm SD}(L)/T]\;,
\]%end{equation}
which allows to estimate $\rho$, the fraction of the total area of
the system covered  by such defects, as
\begin{equation}                          \label{nu}
  \rho(T)\sim \left[\frac{T}{f_{\rm DW}^{(0)}(T)}\right]^2
  \exp\left(-\frac{2E_0}{T}\right)\;.
\end{equation}
By looking when $\rho(T)$ becomes of
the order of 1, a criterium is obtained which differs from Eq.
(\ref{Tc}) only by the replacement $E_{\rm K}\rightarrow E_0$.
Since $E_0\approx E_{\rm K}$, this gives an additional support for
our estimate of the phase transition temperature.

However, in this approach it becomes more clear that the estimate
we have constructed is an estimate from below. Since stripe
defects are strongly anisotropic and can have different
orientations, they have to start crossing each other while $\rho$
is still much smaller than 1. An estimate shows that the average
distance between the centers of stripe defects becomes comparable
with their average length when $T\approx \frac{2}{3}T_c$. Since
each of such crossings costs an additional energy, this will
decrease the rate at which  $\rho(T)$ grows with increasing
temperature [as well as the rate at which $f_{\rm DW}(T)$
decreases].

The defect which is formed when a stripe defect crosses a type I
domain wall with different orientation is shown in Fig. 6(c). Like
the two other types of local defects considered above this defect
is neutral and consists of two clusters of anomalous sizes (four and
two), which suggests that its energy is also close to $E_{\rm K}$.
Since no additional energy scale is involved, one can hope that
the effects related with such crossings will lead only to the
appearance of some numerical factor (comparable with 1) in the
right-hand side of Eq. (\ref{Tc}).

Note, that one also cannot exclude a possibility that the
disordering of vortex pattern is a multistage process and takes
place as a sequence of phase transition, the first of which, at
$T_c\sim 0.01\,J$, is related to the appearance of infinite domain
walls with only one orientation.

At $T\rightarrow 0$ the value of $\rho$ given by Eq. (\ref{nu})
exponentially tends to zero, which means that with the decrease of
temperature the system becomes more and more ordered. Quite
remarkably, this is accompanied by the divergence of the
correlation radius of fluctuations,
\begin{equation}                          \label{rc}
 r_c(T)\approx\frac{T}{2f_{\rm DW}(T)}\;.
\end{equation}
which is determined by the average length of the defect. In the
low temperature limit \makebox{$r_c(T)\propto 1/T$}.

\subsection{Finite-size effects}

Thus, we have found that $T_c$, the temperature of vortex-pattern
ordering in the fully frustrated $XY$ model on a dice lattice
can be expected to be of the order of $0.01\,J$. It has to be
emphasized that at $T\lesssim T_c$ the fluctuation induced free
energy of domain walls is extremely weak, $f^{(0)}_{\rm
DW}\lesssim 10^{-4}\,\gamma$, where $\gamma\ll 0.01$ is an
additional small parameter calculated in Sec. IV.F.

The very low value of the ratio $f^{(0)}_{\rm DW}/T$ at $T\lesssim
T_c$ leads to the unusual prominence of the finite-size effects
consisting in the spontaneous formation of domain walls crossing
the whole system. If a sample has a form of a stripe with a finite
width, $L$, the probability (per unit length) to have a domain
wall crossing the whole system can be estimated as
\[%begin{equation}                             \label{pL}
  p_{\rm}(L)\sim \exp\left[-\frac{f_{\rm DW}(T)}{T}L\right]
  =\exp\left[-\frac{L}{2r_c(T)}\right]\;.
\]%end{equation}
Vortex-pattern ordering, or, at least, any traces of such an
ordering can be expected to be observable only when the average
distance between such walls, $r(L)=1/p(L)$ [for $L\lesssim r_c(T)$
this quantity plays the role of the effective correlation radius
induced by the finite-size effects] is much larger then 1, which
requires to have $L\gg r_c(T)$.

In typical systems with discrete degrees of freedom analyzed in
statistical mechanics (the simplest example being the isotropic
Ising model) a twofold or a threefold decrease of temperature with
respect to $T_c$ is usually sufficient to obtain $r_c\sim 1$,
which allows to observe the ordering even in relatively small
systems. However, in  situations  when a finite free energy of
domain walls arises only from the anharmonic fluctuations, $r_c(T)$
diverges not only when $T\rightarrow T_c$, but also when
$T\rightarrow 0$ \cite{fxk}, and, therefore, the best conditions
for the observation of vortex-pattern ordering in a finite system
are achieved at intermediate temperatures. The differentiation of
$f_{\rm DW}(T)/T$ with respect to $T$ shows that the minimum of
$r_{c}(T)$ defined by Eq. (\ref{rc}) is achieved when
\begin{equation}                             \label{Tm}
  T=\frac{E_{\rm K}}{\ln[\nu E_{\rm K}/f^{(0)}_{\rm DW}(T)]}\;.
\end{equation}
Numerical solution of Eqs. (\ref{Tm}) for the same values of
$E_{\rm K}$ and $\gamma,$ and substitution of the result into Eq.
(\ref{rc}) show that the minimal value of $r_c$ is achieved when
\makebox{$T \approx 0.8\,T_c$} and is of the order of $2\cdot
10^{4}$. Thus, the observation of vortex-pattern ordering requires
the linear size of the system to be at least comparable with
$10^5$.

\section{Destruction of phase coherence}

Up to now we have discussed only one phase transition related to
the disordering of vortex pattern and the proliferation of domain
walls. However, the ground states of uniformly frustrated $XY$
models are characterized by a combination of discrete and
continuous degeneracies, which provides possibilities for the
existence of (at least) two different phase transitions \cite{TJ}.
The second phase transition is related to the vanishing of the
helicity modulus, $\Gamma(T)$, describing the rigidity of the
system with respect to a phase twist. In terms of a Josephson
junction array this phase transition  corresponds to the
destruction of superconductivity. It takes place not necessarily
at the same temperature as the vortex-pattern disordering.

The interaction between the discrete and continuous degrees of freedom
in uniformly frustrated $XY$ models has a non-perturbative nature
and is related to the formation of fractional vortices at corners
and intersections of domain walls \cite{Hals,KU,K86}. According to
the general scheme proposed in Ref. \onlinecite{K86}, three
scenarios are possible in a situation when the disordering of
vortex pattern takes place as a single phase transition (whose
temperature we denote $T_c$) and not as a sequence of phase
transitions \cite{SMK}.

First, the vanishing of the helicity modulus can take place at
$T<T_c$, if $T_{\rm V}$, the temperature of pair dissociation for
ordinary (integer) vortices is lower than $T_c$. The phase
transition at $T=T_{\rm V}$ in that case has exactly the same
nature as the Berezinskii-Kosterlitz-Thouless transition
\cite{BKT} in the conventional $XY$ model (without frustration).
Numerical simulations \cite{MS,ffx-MC}, as well as analysis of
the mutual influence between vortices and kinks on domain walls
\cite{ffx}, demonstrate that this scenario is realized in the fully
frustrated $XY$ models on square and triangular lattices.

The vanishing of the helicity modulus can also take place at
$T>T_c$, but only if at $T=T_c$ the logarithmical interaction of
fractional vortices is strong enough to keep them bound in pairs.
Note that at $T<T_c$ the confinement of fractional vortices is
ensured by their linear interaction related to a finite free
energy (per unit length) of the domain walls which are connecting
them. In such a case the loss of phase coherence is related to the
dissociation of pairs of logarithmically interacting fractional
vortices and can be expected to  take place at $T=T_{\rm FV}>T_c$,
where $T_{\rm FV}$ is the solution of the equation,
\begin{equation}                                \label{univ}
T=\frac{\pi }{2}Q^2\Gamma(T)\;,
\end{equation}
and $Q<1$ is the topological charge of the elementary fractional
vortex. Eq. (\ref{univ}) is nothing else but the generalization of
the Nelson-Kosterlitz universal relation \cite{NK} for fractional
vortices \cite{K86}. This scenario is realized in the
antiferromagnetic $XY$ model on a {\em kagome} lattice, where $T_c$
is expected to be anomalously small, $T_c/J\sim 10^{-4}$ \cite{fxk},
and also in the uniformly frustrated $XY$ model with $f=1/3$ on a
dice lattice, in which the vortex pattern is disordered at any finite
temperature and becomes quasi-ordered only at $T=0$  \cite{xdt}.

The two transitions can be expected to coincide if at $T=T_c$ the
value of $\Gamma(T)$ is sufficiently large to ensure that integer
vortices are bound in pairs, but is not large enough to prevent
from dissociation the pairs of fractional vortices. In such a case
$\Gamma(T)$ jumps to zero exactly at $T=T_c$, the ratio
$T/\Gamma(T)$ at the transition point is not universal \cite{K86}
\makebox{[$(\pi/2)Q^2<T/\Gamma(T)<\pi/2$]}, and the transition is
likely to be of the first order \cite{KU,KLLK}. The results of
numerical simulations suggest that this scenario is quite possibly
realized on square lattice at $f=2/5$ \cite{LT}, as well as at
$f=1/8$ and $f=1/10$ \cite{KLLK}.

In the case of the fully frustrated $XY$ model on a dice lattice
the effective value of the helicity modulus (properly averaged
over angles) at $T=0$ is \makebox{$\Gamma_0 =(2/3)^{3/2}J\approx
0.54\, J$}, % \cite{TTM},
and, therefore, at $T=T_c\ll J$ the integer vortices are strongly
bound in pairs. It has been already mentioned in Sec. V that the
kinks on both types of zero-energy domain walls [on the background
of state (a)] are neutral. Therefore, the mechanism which forces
$T_{\rm V}$ to be lower than $T_c$ in the fully frustrated models
on square and triangular lattices \cite{ffx} here most probably
does not work.

\begin{figure}[b]
\includegraphics[width=65mm]{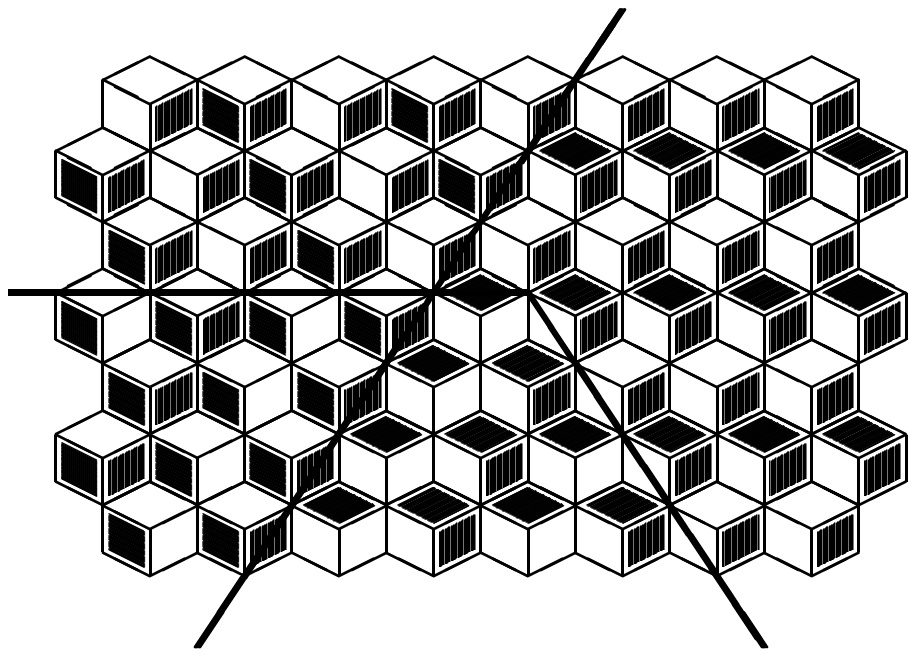}
\caption[Fig. 7] {An example of a fractional vortex}\label{fig7}
\end{figure}

In the considered model the fractional vortices appear at points
where two type I zero-energy domain walls cross each other. Fig. 7
shows an example of such a crossing. Note that  after crossing one
of the walls has to be transformed into a type II domain wall. The
energy of this state is above the ground state energy because one
of vortex clusters contains only two positive vortices instead of
three. The accurate summation of the nominal values of the
variables $\theta_{jk}$ along a closed loop going around this
cluster demonstrates the misfit of $\pi/4$ with respect to what
one could expect from counting the number of positive and negative
half-vortices inside this loop. The value of the misfit is the
same for all closed loops surrounding the anomalous cluster and
should be compensated by a continuous rotation of the phase by the
same angle in the opposite direction. This means that the cluster
consisting of two positive vortices behaves itself as a fractional
vortex whose topological charge is equal to $-1/8$.

One also can construct an analogous intersection where one of the
clusters is formed by four positive vortices instead of three. The
topological charge of such a defect will be equal to $+1/8$. It is
clear that when the cluster of an anomalous size consists of negative
vortices (instead of positive), the sign of the topological charge
is reversed. The topological charges of more complex intersections
(containing, for example, the clusters of five or more vortices or
several clusters of anomalous sizes) will all be the multiples of
$1/8$. Note that the excess vorticity which can  be associated
with a vortex cluster depends not only on its size but also on the
shape. For example, when a cluster consists of three vortices, but
has the shape of a hexagon,  the topological charge which has to
be associated with it is equal to $\pm 3/8$.

The only possibility to have domain walls crossings in an
equilibrium infinite system well below the temperature of
vortex-pattern ordering is related to crossing of stripe defects
discussed in Sec. V.B. Fig. 6(c) shows an example of the
intersection of a stripe defect with a domain wall where the two
fractional vortices have the opposite topological charges, which
makes the energy of such an object finite. Although it is possible
also to construct an example in which the topological charges of
the two intersections will be the same, the total topological
charge of any finite size defect (for example, formed by several
intersecting stripe defects) has to be integer \cite{K86,Hals}.
Therefore, at low temperatures the fractional vortices can be
present only in the form of bound pairs, whose size is restricted
more by the available separations between domain walls in stripe
defects rather then by the logarithmic interaction of these
objects (which is $64$ times weaker than the interaction of
ordinary vortices).

With increase of temperature the average separation between the
domain walls forming a stripe defect increases, which allows
larger separations between the fractional vortices of opposite
sign. Above the temperature of vortex-pattern disordering, the
restrictions for the distances between fractional vortices related
with the separations between domain walls in stripe defects will
no longer exist. It looks rather likely \cite{K86} that in such a
situation one can consider fractional vortices (at the scales that
are large in comparison with the correlation radius) as really
logarithmically interacting objects. The same approach may  be
also applicable to a finite system at $T<T_c$ if its linear size
does not exceed the size-dependent correlation radius $r_c(L)$
introduced in Sec. V.C. In both cases one can speculate about the
possibility of a phase transition related with the unbinding of
neutral pairs formed by logarithmically interacting vortex
clusters of anomalous sizes.

Substitution of $Q=1/8$ into Eq. (\ref{univ}) allows one to find
that the temperature of this phase transition can be estimated as
\[
T_{\rm FV}\approx \frac{\pi}{128} \Gamma_0\approx 0.013\, J\;.
\]
This is an estimate from above which neglects the renormalization
of $\Gamma$ by thermal fluctuations. Comparison with the estimate
$T_c\gtrsim 0.01\,J $ obtained in Sec. VI suggests that in the
fully frustrated $XY$ model on an infinite dice lattice the
destruction of phase coherence will be triggered by the
disordering of vortex pattern, which can be expected to occur at a
temperature where the logarithmic interaction of fractional
vortices at $T_c$ is too weak to keep them bound in pairs.

On the other hand, in a situation when the size of the system is
insufficient to exclude the specific finite-size effects leading
to the disordering of vortex pattern at any temperature (see Sec.
V.C), one still can discuss the possibility of a
phase transition (slightly smeared by the finite-size effects),
in which the loss of phase coherence
will occur as the result of the unbinding of fractional vortices
at $T=T_{\rm FV}\sim 0.01\,J$. In numerical simulations this phase
transition can be observed by analyzing if vortex clusters of
anomalous sizes are bound in neutral pairs or not. However, at
$T\sim 0.01\,J$ one should be specially attentive about checking
if the time of simulation is sufficient for the equilibration of
the vortex subsystem, which at such temperatures will require much
longer times than the equilibration of the spin-wave subsystem.
It is not clear if in numerical simulations of Ref.
\onlinecite{CF} (discussed in a more detail in Sec. VIII) this
condition was really satisfied.

%\newpage
\section{Magnetic effects}

It has been already mentioned in the Introduction that the main
interest to the uniformly frustrated $XY$ models has appeared in
relation to their application  for the description of Josephson
junction arrays in external magnetic field. Since we have found
that in the case of the fully frustrated model on a dice lattice
the order-from-disorder mechanism for the removal of an accidental
degeneracy is extremely inefficient, in a physical situation one
should also take into account  other possible mechanisms. In the
case of a proximity coupled array,  the most important of them is
related to the magnetic interactions of currents
\cite{PH,nwd,footnote4}.

When the proper magnetic fields of currents in the array are taken
into account, the Hamiltonian of the frustrated $XY$ model should
be replaced \cite{SK,DJ} by
\begin{equation}                                      \label{Hm}
H_{\rm}=-J\sum_{(jk)}\cos(\theta_{jk}-a_{jk}) +E_{\rm magn}
%(\{a_{jk}\})
\end{equation}
where $\theta_{jk}\equiv{\varphi_k-\varphi_j-A_{jk}}$ includes
only the contribution from the external magnetic field, defined by
Eq. (\ref{Ajk}), whereas the analogous contribution from the
currents is denoted $a_{jk}$. The second term in Eq. (\ref{Hm}),
\begin{equation}                                    \label{Emagn}
E_{\rm magn}=\frac{1}{2}\sum_{\alpha,\beta}L^{-1}_{\alpha\beta}
\Phi_\alpha \Phi_\beta\;,
\end{equation}
is the energy of the current induced (screening) magnetic fields
expressed in terms of
\begin{equation}
\Phi_\alpha=\frac{\Phi_0}{2\pi} \left.\sum_{\Box} a_{jk}\right.\;,
\end{equation}
the magnetic fluxes of these fields through the plaquettes of the
array (denoted by Greek letters), $L_{\alpha\beta}$ being the
matrix of mutual inductances \cite{DK} between the plaquettes.

The values of the variables $a_{jk}$ should be found from
the minimization of $H$. The result  of the variation of Eq.
(\ref{Hm}) with respect to $a_{jk}$ can be rewritten as
\begin{equation}                             \label{Phial}
\Phi_\alpha=\sum_\beta L_{\alpha\beta}I_\beta\;,
\end{equation}
where $I_\beta$ is so-called mesh current  \cite{DK,PZWO} which
can be associated with the plaquette $\beta$. The currents in the
junctions,
\[%begin{equation}                             \label{Ijkm}
I_{jk}=I_0\sin(\theta_{jk}-a_{jk})\;,
\]%end{equation}
are given by the difference of mesh currents in the two
plaquettes, $\alpha_{jk}$ and $\alpha'_{jk}$, sharing the bond
$(jk)$,
\begin{equation}                            \label{Ijkmm}
I_{jk}=I_{\alpha_{jk}}-I_{\alpha'_{jk}}\;.
\end{equation}
The substitution of Eq. (\ref{Phial}) into Eq. (\ref{Emagn})
allows to rewrite it in the more familiar form  as
\[%begin{equation}                             \label{Emagn2}
E_{\rm
magn}=\frac{1}{2}\sum_{\alpha,\beta}{L_{\alpha\beta}}I_\alpha
I_\beta\;.
\]%end{equation}

In the regime of weak screening one can assume that the values of
the currents are not affected by their magnetic fields and
calculate $E_{\rm magn}$ replacing $I_{jk}$ by
\makebox{$I^{(0)}_{jk}=I_{0}\sin\theta_{jk}$.} Nonetheless, the
calculation of the first term in Eq. (\ref{Hm}) with the same
accuracy requires to take into account its dependence of $a_{jk}$.
The contribution to this term which is linear in $a_{jk}$ has a
form
\[
-\sum_{(jk)}I^{(0)}_{jk}a_{jk}\;,
\]
and with the help of Eqs. (\ref{Emagn})-(\ref{Ijkmm}) can be shown
to be equal to $-2E_{\rm magn}$, where $E_{\rm magn}$ is
calculated for the ``bare" values of currents, $I^{(0)}_{jk}$.
Therefore, in the regime of weak screening, $H_{\rm magn}$, the
total magnetic correction to the Hamiltonian of a Josephson
junction array is equal to $E_{\rm magn}$, but has the opposite
sign,
\[H_{\rm magn}=-E_{\rm magn}\;.
\]

The comparison of $E_{\rm magn}$ in different periodic states in a
fully frustrated superconducting wire network with the dice
lattice geometry has been recently made in Ref. \onlinecite{nwd}.
Since in terms of the gauge-invariant phase differences
$\theta_{jk}$ the structure of these states is in one-to-one
correspondence with the ground states of the fully frustrated $XY$
model with the same geometry, the same results are also applicable
to an array formed by weakly coupled superconducting islands. It
follows from the results of Ref. \onlinecite{nwd} that in arrays
the expression for $E_{\rm magn}$ (normalized per a sixfold
coordinated site) in the regime of weak screening can be written
as
\begin{equation}                             \label{Emagn3}
E_{\rm magn}=\frac{\epsilon}{3}\frac{J^2}{E_\Phi}\;,
\end{equation}
where
\[
E_\Phi=\frac{\Phi_0^2}{4\pi^2a}
\]
is the characteristic energy scale which depends on $a$, the lattice
constant of the array. The
numerical coefficient $\epsilon$ in Eq. (\ref{Emagn3}) is
determined by the structure of vortex pattern in the considered
ground state and can be expressed as a linear combination of the
coefficients $\lambda_i>0$ defining the mutual inductances,
$L_i\equiv -\lambda_i a $,  between different pairs of plaquettes
on a dice lattice. Here $i=1$ corresponds to the nearest neighbors,
$i=2$ to the next-to-nearest neighbors, {\em etc.}.

The main conclusion of Ref. \onlinecite{nwd} is that in the fully
frustrated system with the dice lattice geometry the coefficient
$\epsilon$ is the largest in the state (c). For example, the
numerical coefficient in the expression,
$$
\delta H^{\rm}_{\rm magn}=H^{\rm a}_{\rm magn}-H_{\rm magn}^{\rm
c}=\mu^{}\frac{J^2}{E_\Phi}\;,
$$
for the difference between the magnetic energies of the states (a)
and (c),
\begin{eqnarray*}
\mu^{\rm}= \frac{\epsilon^{\rm c}-\epsilon^{\rm a}}{3} & = &
\frac{1}{6}\left(\lambda_2+3\lambda_3-\lambda_4-6\lambda_5\right. \\
& & \left.-5\lambda_7+3\lambda_8+4\lambda_{10}-\ldots\right)
\approx 0.17\;,
\end{eqnarray*}
is positive.

For $a=8\,\mu\mbox{m}$ \cite{TTM} $E_\Phi\approx 0.98\cdot
10^4\,\mbox{K}$, which shows that at $T\sim J$ the differences
between the magnetic energies are extremely small, $\delta H_{\rm
magn}/T\lesssim 10^{-4}$, and are even smaller than the
differences between the free energies of anharmonic fluctuations.
In this estimate we have assumed $T\lesssim 10\,\mbox{K}$.

It should be emphasized that in proximity coupled arrays arrays
the coupling constant $J$ has a strong temperature dependence in a
wide interval of temperatures, so the decrease of the
dimensionless temperature $\tau=T/J(T)$ with the decrease of real
temperature $T$ is much more strongly influenced by the growth of
$J(T)$ than by the decrease of $T$. Roughly speaking, in the
experiments of Ref. \onlinecite{TTM} the decrease of $T$ by $1\;$K
corresponds to the decrease of $\tau$ by one order of magnitude.

This suggests that the importance of magnetic effects rapidly
grows with the decrease of $\tau$. Comparison of $\delta H^{\rm
}_{\rm magn}$ with $\delta F^{\rm }_{\rm anh}=\gamma T^2/J$ shows
that the magnetic energy becomes dominant when
\[
\tau<\tau_{\rm magn}=\left(\frac{\mu^{\rm }}{\gamma^{\rm
}}\frac{T}{E_\Phi}\right)^{1/3}\approx 0.30\;,
\]
where we have put $T=5\,\mbox{K}$. Thus at $\tau\lesssim 0.1$,
where the vortex-pattern ordering can be expected to occur, one
can take into account only the magnetic energies of different
states (or domain walls), completely neglecting the free energy of
anharmonic fluctuations. Therefore, the structure of the periodic
vortex pattern in the low temperature phase of a proximity coupled
array should be of the type (c).

The energy of domain walls separating different versions of state
(c) from each other, $E_{\rm DW}$, will be close to $\delta H_{\rm
magn}$, which will make the vortex pattern more robust with
respect to thermal fluctuations in comparison with the $XY$ model.
However, any quantitative conclusions about the temperature of the
phase transition(s) related to vortex-pattern disordering in this
situation are impossible without the detailed analysis of the
energies and other properties of the topological excitations in
state (c) [analogous to the one performed in Sec. V for state
(a)], which goes beyond the limits of this article. Nonetheless,
some increase of the region of stability of the low-temperature
phase with the ordered vortex pattern is inevitable, which allows
to conclude that the scenario in which the unbinding of fractional
vortices with topological charges $\pm 1/8$ takes place as an
independent phase transition is impossible.

Inclusion of the magnetic interactions into analysis improves also
the situation with respect to the finite-size effects. For
\makebox{$E_{\rm DW}\sim J^2/E_\Phi$} the probability to have a
domain wall crossing the whole system (of the width $L$) becomes
negligible for
\[
L\gg L_c\sim \frac{E_\Phi}{T}\tau^2\;,
\]
which at $\tau\ll 1$ is a much more mild restriction than the one
obtained in Sec. V for the $XY$ model.

On the other hand, it is known that the magnetic interactions of
currents, or, in other words, screening effects, lead to the
increase of barriers a vortex has to overcome when moving between
the plaquettes of an array \cite{PZWO}. Thus, although the growth
of screening effects with a decrease in temperature improves the
stability of a vortex-pattern ordering, it simultaneously leads to
the further increase of (already exponentially large) times
required  for the relaxation of the system to equilibrium. This
may be one of the reasons why the finite-frequency experiments of
Ref. \onlinecite{TTM} have not demonstrated any signs of a phase
transition at $f=1/2$.

It should also be noted that in the case when the loss of phase
coherence is related to dissociation of fractional-vortex pairs,
the universal relation \cite{BMO} for the value of the
two-dimensional magnetic penetration depth \cite{Lambda},
$\Lambda$, at the transition temperature can be rewritten (in our
notation) as
\[
\frac{\Lambda(T_{\rm FV})}{a}=\frac{Q^2}{8}\frac{E_{\Phi}}{T_{\rm
FV}}\;,
\]
As a consequence, the smearing of the phase transition due to the
finiteness of $\Lambda$ in the case when it is related to
unbinding of fractional vortices should be more pronounced then in
the case of integer vortices.

\section{Conclusion}

In the present work we have investigated the fully frustrated $XY$
model on a dice lattice and have demonstrated that the energy of
this system is minimized in the highly degenerate family of
states (described in Sec. II.C), in which the vortices of the same
sign are grouped into clusters of three.
The accidental degeneracy of these states can be described in terms
of the formation of a network of intersecting zero-energy domain
walls, whereas the residual entropy related to this degeneracy is
proportional to the linear size of the system,
as in the case of a honeycomb lattice \cite{K86}.

The central result of this article consists in determining the
structure of the periodic vortex pattern which is selected
at low temperatures by thermal fluctuations. It is shown in Fig.
3(a). However, this effect is rather weak, being induced only by
the anharmonic fluctuations. As a consequence of a hidden gauge
symmetry, the free energy of the harmonic fluctuations turns out
to be the same for all ground states. The same conclusion is
applicable also to quantum generalizations of the model both at a
finite and at zero temperature, when one should speak of
zero-point fluctuations.

The destruction of the periodic vortex pattern  with the increase
of temperature is related to the proliferation of domain walls.
The dimensionless temperature, \makebox{$\tau=T/J$}, at which the
corresponding phase transition can be expected to take place can
be estimated as $\tau_c\sim 0.01$, where one factor of $0.1$ is
related to the smallness of the energy of  particular point-like
defects on domain walls, and the other (which is of the
logarithmic origin) to the extreme smallness of the fluctuation
induced free energy of zero-energy domains walls. The analysis of
possible scenarios suggests that the loss of phase coherence in
the considered system can be expected to take place at the same
temperature as the disordering of vortex pattern.

The extreme smallness of the fluctuation induced free energy of
domain walls will manifest itself also in the huge prominence of
the finite-size effects consisting in the appearance of domain
walls crossing the whole system and leading to the disordering of
vortex pattern even at $\tau<\tau_c$. As a consequence, even at
the ``optimal" temperature, $\tau\approx 0.8 \tau_c$, the linear
size of the system required to observe the vortex-pattern ordering
should be much larger than $r_c^{\rm min}\approx2\cdot 10^4$.
However, in smaller systems one can still discuss a possibility
for the observation of a phase transition related to the loss of
phase coherence and consisting in unbinding of pairs of
logarithmically interacting vortex clusters of anomalous sizes,
which can be expected to happen at $\tau_{\rm FV}\sim 0.01$.

Our conclusions are consistent with the results of the {numerical
simulations} of the same model by Cataudella and Fazio \cite{CF},
who have investigated the temperatures down to $\tau=0.01$ and
have found no signs of vortex-pattern ordering. The results
presented in Sec. V demonstrate that even if the lowest
temperatures investigated in Ref. \onlinecite{CF} were indeed
below $\tau_c$, the size of the system was definitely not
sufficient for the observation of such an ordering. In the
notation based on counting only the sixfold coordinated sites
(used in this work), the largest size of the system analyzed
in Ref. \onlinecite{CF}
corresponds to $56\times 42$, whereas the majority of the data has
been taken at $24\times 18$. It seems rather likely that analogous
simulations of a $10^5\times 10^5$ system, which may be
required for the observation of vortex-pattern ordering in the
fully frustrated $XY$ model on a dice lattice, can become possible
only with a further development of computational abilities.

Another result of the numerical simulations of Ref.
\onlinecite{CF} is related to the dynamical properties of the
system (which were not discussed in this article) and consists in
finding below $\tau_*\approx 0.06$ the signs of a {glassy
behavior}. Namely, the relaxation of the total energy at
$\tau<\tau_*$ becomes logarithmic in time in contrast to the
exponential relaxation at $\tau>\tau_*$, whereas the behavior of
the autocorrelation function of the variables $\theta_{jk}$
demonstrates a dependence on the waiting time.

One can certainly make an attempt to relate this observation to
the specific features of the considered system.
For the temperatures and system sizes analyzed in Ref.
\onlinecite{CF} one can safely neglect the fluctuation induced
free energy of zero-energy domain walls and treat them as objects
with zero free energy. It seems rather likely then that the
typical state obtained after cooling down the system from
$\tau\sim 1$ to $\tau\ll E_K/J$ can be described as  a network
formed by zero-energy domain walls, which contains a large number
of point-like defects, such as kinks and intersections, which cost
an additional energy. We know that the intersections with zero
energy are also possible [see Fig. 3(f)], but one can expect that
the majority of the intersections formed during cooling down from
a disordered state will not have the optimal structure necessary
for that.

The glassy behavior can be then expected from the necessity of the
disentanglement of this domain wall network with the decrease of
temperature. For different temperatures above $\tau_c$, the
equilibrium concentration of point-like defects should be
different. However, in contrast to kinks on a domain wall whose
number can be changed due to annihilation of two kinks of opposite
signs (which may be a relatively fast process), to change the
number of the intersections of domain walls one has to change the
number of these walls, which will require much longer times than
the annihilation of kinks. In particular, it seems rather likely
that the processes related to the annihilation of zero-energy
domain walls crossing the whole system will be characterized by
relaxation times diverging with the system size, providing thus a
source for a genuine glass-like behavior.

On the other hand, even if the times characterizing the relaxation
of vortex pattern are not divergent in the thermodynamic limit and
are restricted only by the barriers whose typical height remains
of the order of $J$, this already can be the source for a
glassy-like behavior (associated with a wide distribution of such
barriers) at $\tau\ll 1$ in a wide interval interval of times. For
example, $\exp(1/0.05)\sim 10^9$, whereas in Ref. \onlinecite{CF}
the anomalous relaxation has been studied only at much shorter
times. In such a case one can argue that the observation of a
glassy-like behavior in the considered $XY$ model is possible due
to a combination of three factors, namely, (i) the existence of
zero-energy domain walls, (ii) the special ineffectiveness of the
order-from-disorder mechanism for the removal of an accidental
degeneracy and (iii) the anomalously low transition temperature
($\tau_c\sim 0.01$). However, the choice between the two scenarios
(genuine glass vs. the dynamical freezing of vortex relaxation)
should be made on the basis of a much more detailed analysis of
the domain-walls disentanglement.

In the experimental situation, the magnetic interactions of
currents in a Josephson junction array will be of greater
importance for the stabilization of a particular vortex pattern
then the anharmonic fluctuations. This mechanism leads to the
selection of the pattern shown in Fig. 3(c) and makes the periodic
vortex pattern less vulnerable with respect to fluctuations.

\acknowledgments

The author is grateful to B. Dou\c{c}ot for numerous useful
discussions. This work has been supported in part by the Program
"Quantum Macrophysics" of the Russian Academy of Sciences, by the
Program "Scientific Schools of the Russian Federation" (grant No.
1715.2003.2) and  by the Swiss National Science Foundation.

%\newpage $~$


\begin{thebibliography}{99}
\bibitem{fn1} Although in Sec. III quantum
generalizations of the same model are also considered, the term
``Hamiltonian" in this article is always used as in statistical
mechanics, to denote the expression for the static energy of
the system, and not as in quantum mechanics.
\bibitem{footnote3}{In all examples shown in Fig. 3 all domain walls
               of the type I (if present) are horizontal.}
\bibitem{footnote4} In the case of SIS
(superconductor-insulator-superconductor) array, it is also
necessary to consider another mechanism of the removal of an
accidental degeneracy related to the anharmonic quantum
fluctuations.
\bibitem{Lambda} This is the length at which the logarithmical interaction
                 of vortices is screened \cite{SK,Pearl}.
\end{thebibliography}
\end{document}